\documentclass[nofootinbib,showpacs,superscriptaddress,fleqn]{iopart}

\usepackage{amsfonts}
\usepackage{amstext}
\usepackage{amssymb}
\usepackage{bm}
\usepackage{cite}
\usepackage{dcolumn}
\usepackage{graphicx}
\usepackage{graphics}
\usepackage[latin1]{inputenc}
\usepackage{latexsym}
\usepackage{rotating}
\usepackage{url}
\usepackage[colorlinks=true]{hyperref}
\usepackage{xspace} % Sensible space treatment at end of simple macros
\usepackage[usenames]{color}
\usepackage{booktabs}
\usepackage{graphicx}    
\usepackage{grffile}

\newcommand{\be}{\begin{equation}}
\newcommand{\ee}{\end{equation}}
\newcommand{\Inte}{{\mbox{\tiny int}}}
\newcommand{\dCS}{{\mbox{\tiny dCS}}}
\newcommand{\Kerr}{{\mbox{\tiny Kerr}}}
\newcommand{\SRKerr}{{\mbox{\tiny SR-Kerr}}}
\newcommand{\resum}{{\mbox{\tiny resum}}}
\newcommand{\expp}{{\mbox{\tiny exp}}}
\newcommand{\eff}{{\mbox{\tiny eff}}}
\newcommand{\inv}{{\mbox{\tiny inv}}}
\newcommand{\GR}{{\mbox{\tiny GR}}}
\newcommand{\HT}{{\mbox{\tiny HT}}}

\newcommand{\nn}{\nonumber}

\definecolor {darkgreen}{rgb}{0.2,0.7,0.2}

\begin{document}

\title[The exact dCS metric for a spinning BH possesses a fourth constant of motion: ...]{The exact dynamical Chern Simons metric for a spinning black hole possesses a fourth constant of motion: A Dynamical-Systems-Based Conjecture}

\author{%
Alejandro~C\'ardenas-Avenda\~no$^{1}$ $^{2}$, 
Andr\'es F. Gutierrez$^{3}$,  Leonardo A. Pach\'on$^{3}$
and
Nicol\'as~Yunes$^{1}$
}

\address{$^{1}$~eXtreme Gravity Institute, Department of Physics, Montana State University, Bozeman, MT 59717, USA.}

\address{$^{2}$~Programa de Matem\'atica; Fundaci\'on Universitaria Konrad Lorenz, 110231 Bogot\'a, Colombia}

\address{$^{3}$~Grupo de F\'isica Te\'orica y Matem\'atica Aplicada, Instituto de F\'{\i}sica, Facultad de Ciencias Exactas y Naturales, Universidad de Antioquia, Calle 70 No. 52-21, Medell\'in, Colombia.}

\date{\today}

\begin{abstract}

The recent gravitational wave observations by the LIGO/Virgo collaboration have allowed the first tests of General Relativity in the extreme gravity regime, when comparable-mass black holes and neutron stars collide. 
Future space-based detectors, such as the Laser Interferometer Space Antenna, will allow tests of Einstein's theory with gravitational waves emitted when a small black hole falls into a supermassive one in an extreme mass-ratio inspiral. 
One particular test that is tailor-made for such inspirals is the search for chaos in extreme gravity. 
We here study whether chaos is present in the motion of test particles around spinning black holes of parity-violating modified gravity, focusing in particular on dynamical Chern-Simons gravity. 
We develop a resummation strategy that restores all spin terms in the General Relativity limit, while retaining up to fifth-order-in-spin terms in the dynamical Chern-Simons corrections to the Kerr metric. 
We then calculate Poincar\'e surfaces of section and rotation numbers of a wide family of geodesics of this resummed metric. 
We find evidence for geodesic chaos, portrayed by thin chaotic layers surrounded by deformed invariant tori. This chaotic layers shrink in size as terms of higher-order in spin are included in the dynamical Chern-Simons corrections to the Kerr metric.
Our numerical findings suggest that the geodesics of the as-of-yet unknown exact solution for spinning black holes in this theory may be integrable, and that there may thus exist a fourth integral of motion associated with this exact solution.
The studies presented here begin to lay the foundations for chaotic tests of General Relativity with the observation of extreme mass ratio inspirals with the Laser Interferometer Space Antenna.

\end{abstract}

\submitto{\CQG}

\maketitle

%%%%%%%%%%%%%%%%%%%%%%%%%%%%%%%%%%%%%%%%%%%%%
%%%%%%%%%%%%%%%%%%%%%%%%%%%%%%%%%%%%%%%%%%%%%

%%%%%%%%%%%%%%%%%%%%%%%%%%%%%%%%%%%%%%%%%%%%%
%%%%%%%%%%%%%%%%%%%%%%%%%%%%%%%%%%%%%%%%%%%%%
\section{Introduction}
\label{Introduction}

The recent gravitational wave (GW) observations by the LIGO/Virgo collaboration have allowed the first tests of General Relativity (GR) in the extreme gravity regime, when comparable-mass black holes (BH) and neutron stars collide~\cite{Abbott:2016blz,Yunes:2016jcc,TheLIGOScientific:2017qsa,Yunes:2013dva}. Future space-based detectors, such as Laser Interferometer Space Antenna (LISA)~\cite{danzmann2016laser}, will allow tests of Einstein's theory with GWs emitted when a small BH falls into a supermassive (SMBH) one in an extreme mass-ratio inspiral (EMRI)~\cite{AmaroSeoane:2010zy,AmaroSeoane:2012tx}. The latter allow for particularly interesting tests because the inspiral is very sensitive to the SMBH's geometry, the loss of the binary's energy and angular momentum, and non-linear self-force effects.   

One particular probe of extreme gravity that is tailor-made for EMRI signals relates to chaos. For Hamiltonian systems, chaos refers to the non-integrability of the equations of motion, i.e.~the non-existence of a smooth analytic function that interpolates between orbits, and has nothing to do with a system being non-deterministic~\cite{Levin:2006zv}. Being a non-linear theory, one may expect chaos to develop in GR. Nevertheless, EMRIs in GR can be approximated, to zeroth-order, as geodesics of the Kerr spacetime, and the latter has enough symmetries to guarantee that geodesics are completely integrable and thus not-chaotic. Although chaos has been found in the inspiral of spinning BHs of comparable-mass~\cite{Levin:1999zx,Cornish:2003uq,Gopakumar:2005zz}, these features are damped away by gravitational wave dissipation~\cite{Cornish:2003uq}. The presence of large chaotic features in the GWs emitted by EMRIs could then signal either a departure from the strong-equivalence principle or a violation of the Kerr hypothesis\footnote{Small chaotic features could arise due to self-force effects in GR, but in EMRIs such features would be suppressed by the mass ratio, and they would be damped away by dissipative effects, probably rendering them unobservable by LISA.}.
 
What are the signatures of chaos in the GWs emitted by EMRIs? When geodesics are chaotic, the orbital phase space breaks up into islands of instability and prolonged resonant regimes that cause abrupt and large changes in the fundamental frequencies of the motion~\cite{Kiuchi:2004bv,LukesGerakopoulos:2010rc}. These frequencies are encoded directly in the GWs emitted, presumably allowing us to search for them with LISA observations. Abrupt jumps in the fundamental frequencies do also occur in GR, as recently found in the small mass-ratio expansion of the self-force~\cite{Flanagan:2010cd,Gair:2010iv}. These changes, however, are expected to be smaller than those that could arise due to chaos. Therefore, if an EMRI signal is detected with LISA and no chaotic features are detected, then one could place constraints on deviations from the strong-equivalence principle and the Kerr hypothesis.

The development of this idea is still in its infancy. The first step would be to find and study a modified gravity theory in which EMRIs are chaotic. But even if we approximate EMRIs as a sequence of geodesics of a given non-Kerr spacetime, finding whether these geodesics are chaotic is not a trivial task. One approach is to find as many integrals of motion as there are degrees of freedom in the problem, e.g. through Hamilton-Jacobi theory, Painlev\'e analysis~\cite{contopoulos2013order}, Lax pairs~\cite{Cariglia:2014ysa} or Killing tensors~\cite{sommers1973killing}. In GR, this program was completed in the late 1960s and 1970s when Carter found a fourth integral of the motion for the Kerr spacetime~\cite{Carter:1968rr}, and Walker and Penrose showed that the existence of the constant follows from the existence of a symmetric Killing tensor~\cite{Walker:1970un}.

But the analytic approach is not well-suited to all problems. Spinning BH solutions in modified gravity theories have typically only been found in the slow-rotation and small-coupling approximations, i.e.~assuming the dimensionless spin parameter is much smaller than unity and the BH is a small deformation away from the Kerr metric. This approximations make it hard, if not impossible, to find exact integrals of the motion. Such is the case, for example, in dynamical Chern-Simons (dCS) gravity, an effective theory that introduces parity violating interactions to the Einstein-Hilbert action~\cite{jackiw:2003:cmo,Alexander:2009tp}. Spinning black hole solutions have been found in dCS gravity but only in the slow-rotation approximation~\cite{Yunes:2009hc,Yagi:2012ya,Maselli:2017kic}, for which a fourth integral of the motion has not been found from a rank two, Killing tensor~\cite{Yagi:2012ya}. One cannot, however, rule out the existence of a higher-rank Killing tensor, and thus, one cannot make generic statements about chaos in dCS EMRIs.

When analytic techniques fail, one can employ a numerical approach to search for chaos. The idea here is to study the phase space of the system and search for certain chaotic features, e.g., chaotic layers or Birkhoff chains of islands. These features can be found through the study of Poincar\'e surface of sections and the rotation number~\cite{laskar1993frequency}, chaotic scattering~\cite{Dettmann:1994dj}, topological entropies~\cite{Cornish:1996de} or Lyapunov exponents~\cite{contopoulos2013order}. If no features are found at a given resolution, one cannot necessarily conclude that there is no chaos in the system, as chaotic features could be hiding at scales smaller than the numerical resolution employed. Therefore, numerical methods require an extensive study of the parameter space at various resolutions to ensure that no chaotic features are missed. 

Are the geodesics of a spinning BH in dCS gravity, and therefore EMRIs in this theory, chaotic or not? This is the question we tackle in this paper through an extensive and comprehensive numerical analysis. In particular, we calculate and study the Poincar\'{e} surfaces of section and rotation numbers for a wide family of geodesics. We begin by considering the problem of geodesics of the Kerr spacetime, which we know are integrable. Our numerical analysis confirms this expectation when using the exact Kerr metric, but if one employs a slow-rotation expansion of the Kerr background, then clear chaotic features arise. We verify that these features shrink in size when the spin parameter is decreased and when the order of the truncation of the slow-rotation expansion is increased. 

%Result 2 (dCS)
With this new understanding at hand, we then consider geodesics of spinning BHs in dCS gravity. We use both a slow-rotation expansion of the dCS metric, as well as a resummation we develop in this paper, which properly accounts for all spin terms in the GR limit and terms up to fifth order in spin in the dCS correction. When using the resummation, we find no evidence of chaos and only deformations of the invariant tori, proving that geodesics of this metric are slightly non-integrable. Moreover, we find that these deformations shrink as higher order in spin corrections are included in the dCS correction to the Kerr metric. This suggests that if all spin terms of the dCS correction were included in the background spacetime, then geodesics would be exactly integrable, requiring the existence of a (yet-to-be-found) fourth integral of the motion. Given previous results that prove the non-existence of a second-rank Killing tensor for the slowly-rotating dCS metric~\cite{Yagi:2012ya}, our results suggest the existence of a higher-than-second rank Killing tensor associated with the exact dCS metric valid to all orders in spin.

The remainder of this paper presents the details of the results summarized above and it is organized as follows. 
Section~\ref{IABC} reviews briefly dCS gravity, while Sec.~\ref{Framework} summarizes the main methods we use to study chaos. 
Section~\ref{Geodesics} studies geodesics in the Kerr metric and its slow-rotation expansion, while Sec.~\ref{GeodesicsdCS} repeats the analysis in dCS gravity.
Section~\ref{disc} concludes and points to future work. 
~\ref{app:ExplicitMetric} provides explicit expressions for the dCS corrections to the Kerr metric.
Throughout the paper, we use geometric units in which $G=1=c$, and the $(-,+,+,+)$ metric signature. In all figures we set all masses to $1$.

%%%%%%%%%%%%%%%%%%%%%%%%%%%%%%%%%%%%%%%%%%%%%
%%%%%%%%%%%%%%%%%%%%%%%%%%%%%%%%%%%%%%%%%%%%%
\section{BHs in Dynamical Chern Simons Gravity}
\label{IABC}
%%%%%%%%%%%%%%%%%%%%%%%%%%%%%%%%%%%%%%%%%%%%%
%%%%%%%%%%%%%%%%%%%%%%%%%%%%%%%%%%%%%%%%%%%%%

In this section, we briefly describe dCS gravity and its BH solutions, while establishing notation -- for further details see, for instance, Ref.~\cite{Yunes:2009hc}. DCS gravity is a four-dimensional, effective theory that derives from loop quantum gravity~\cite{Taveras:2008yf}, string theory~\cite{Alexander:2004xd} and inflation~\cite{Weinberg:2008hq}. The theory is defined through the action~\cite{Yunes:2009hc}
\begin{eqnarray}
S &\equiv & \int d^4x \sqrt{-g} \Big\{ \kappa_g R + \frac{\alpha}{4} \vartheta R_{\nu\mu \rho \sigma} {}^* R^{\mu\nu\rho\sigma} \nonumber \\ 
& & - \frac{\beta}{2} [\nabla_\mu \vartheta \nabla^{\mu} \vartheta + 2 V(\vartheta) ] + \mathcal{L}_\mathrm{mat} \Big\}\,,
\label{actionCS}
\end{eqnarray}
where, $g$ is the determinant of the metric $g_{\mu\nu}$, $\kappa_g \equiv (16\pi)^{-1}$, $\alpha$ and $\beta$ are coupling constants, $R_{\mu\nu \delta \sigma}$ is the Riemann tensor, and ${}^* R^{\mu\nu\rho\sigma}$ is the dual of the Riemann tensor, defined by
\begin{eqnarray}
{}^* R^{\mu\nu\rho\sigma} \equiv \frac{1}{2} \epsilon^{\rho\sigma\alpha\beta} R^{\mu\nu}{}_{\alpha\beta}\,,
\end{eqnarray}
with $\epsilon^{\mu\nu\alpha\beta}$ the Levi-Civita tensor. The quantity $\vartheta$ is a pseudo-scalar field with potential $V(\vartheta)$, although here we set the potential to zero since the scalar must be massless to preserve shift symmetry. The quantity $\mathcal{L}_\mathrm{mat}$ is the matter Lagrangian density, which couples directly to the metric tensor only. One can check that this theory is diffeomorphism invariant, although the Birkhoff theorem~\cite{Grumiller:2007rv} and the effacement principle~\cite{Yagi:2011xp} are violated. The pseudo-scalar field $\vartheta$ and the coupling constant $\beta$ are taken to be dimensionless, while $\alpha$ has dimensions of length squared. Deformations from GR are proportional to the following dimensionless coupling parameter 
\begin{equation}
\zeta \equiv \frac{\alpha^2}{\kappa_g \beta {\cal{M}}^4}\,,
\label{zetaparameter}
\end{equation}
where ${\cal{M}}$ is a characteristic length of the system; for EMRIs, this length scale is the mass of the SMBH ${\cal{M}} = M$. 

Despite the several and extensive studies of BHs in this theory over the past decade, only approximate solutions have been found in the slow-rotation limit~\cite{Yunes:2009hc,Yagi:2012ya,Maselli:2017kic} and in the extremal limit~\cite{mcnees2016extremal}, always assuming a small-coupling expansion. The slow-rotation approximation assumes that the BH spin angular momentum $S$ is much smaller than its mass squared $\chi= S/M^{2} \ll 1$. The small-coupling expansion postulates that the deformation away from GR is small, which corresponds to a dCS dimensionless coupling constant much smaller than unity $\zeta \ll 1$. In this paper, we study geodesic motion using the BH solutions found in Refs.~\cite{Yunes:2009hc,Yagi:2012ya,Maselli:2017kic}, which were derived using perturbation theory techniques and are valid to fifth order in the spin and to first order in the coupling parameter. 

The slow-rotation and the small-coupling approximations may introduce spurious features in the solution that would not appear in the exact solution, and thus, one typically resums the approximation in an attempt to minimize these features. By resummation, we here mean the introduction of higher order terms in $\chi$ that have not yet been calculated but that one suspects should be present in the solution. For example, a choice of resummation is to replace all $\zeta$-independent terms in the solution with the exact Kerr metric. In a broad sense, there are infinitely many ways to resum the metric and, since the exact solution is unknown, there is no way ensure the chosen resummation is correct. Nonetheless, one expects that for small $\chi$ and $\zeta$, the results obtained should be independent of the resummation\footnote{An analysis that compared the Kerr spectrum to a spectrum from a slowly rotating expansion of Kerr suggested that the choice of resummation described above may be accurate up to extremely high spins $\chi \approx0.9$~\cite{Ayzenberg:2016ynm}.}. 

The slow-rotation and small-coupling corrections to the Kerr metric in dCS gravity is given by the line element~\cite{Yagi:2011xp}
\begin{equation}
\delta(ds^{2})_{\dCS}= 2 g_{t\phi}^\dCS dtd\phi + g_{tt}^\dCS dt^2 + g_{rr}^\dCS dr^2 + g_{\theta \theta}^\dCS d\theta^2 + g_{\phi\phi}^\dCS d\phi^2
\label{solution}
\end{equation}
where $(t,r,\theta,\phi)$ are Boyer-Lindquist coordinates and 
\begin{eqnarray}
g_{tt}^\dCS &=& \zeta \chi^2 \frac{M^3}{r^3} \Bigg[  \frac{201}{1792} \left( 1+\frac{M}{r} +\frac{4474}{4221} \frac{M^2}{r^2} -\frac{2060}{469} \frac{M^3}{r^3} +\frac{1500}{469} \frac{M^4}{r^4} \right. \nonumber \\ 
& & \left. - \frac{2140}{201} \frac{M^5}{r^5}  + \frac{9256}{201} \frac{M^6}{r^6} - \frac{5376}{67} \frac{M^7}{r^7}  \right) (3\cos^2 \theta -1)  \nn \\ & & - \frac{5}{384} \frac{M^2}{r^2} \left( 1 + 100 \frac{M}{r} + 194\frac{M^2}{r^2} + \frac{2220}{7} \frac{M^3}{r^3} \right. \nonumber \\ & & \left. - \frac{1512}{5} \frac{M^4}{r^4} \right) \Bigg] \,, \\
g_{rr}^\dCS &=&   \zeta \chi^2 \frac{M^3}{r^3 f(r)^2} \Bigg[  \frac{201}{1792}  f(r) \left( 1+ \frac{1459}{603} \frac{M}{r} +\frac{20000}{4221} \frac{M^2}{r^2} +\frac{51580}{1407} \frac{M^3}{r^3} \right. \nonumber \\ 
& & \left. -\frac{7580}{201} \frac{M^4}{r^4} - \frac{22492}{201} \frac{M^5}{r^5}  - \frac{40320}{67} \frac{M^6}{r^6} \right) (3 \cos^2 \theta -1)  \nonumber \\ 
& & - \frac{25}{384} \frac{M}{r} \left( 1 + 3\frac{M}{r} + \frac{322}{5} \frac{M^2}{r^2} \right. \nonumber \\ & & \left. + \frac{198}{5} \frac{M^3}{r^3} + \frac{6276}{175} \frac{M^4}{r^4} - \frac{17496}{25} \frac{M^5}{r^5}   \right) \Bigg] \,,\\
g_{\theta\theta}^\dCS &=& \frac{201}{1792} \zeta \chi^2 M^2 \frac{M}{r} \left( 1 + \frac{1420}{603} \frac{M}{r} + \frac{18908}{4221} \frac{M^2}{r^2} + \frac{1480}{603} \frac{M^3}{r^3} \right. \nonumber \\ & & \left. + \frac{22460}{1407} \frac{M^4}{r^4} + \frac{3848}{201} \frac{M^5}{r^5} + \frac{5376}{67} \frac{M^6}{r^6} \right) (3 \cos^2 \theta -1)\,, 
\\
g_{\phi\phi}^\dCS &=&  \sin^2 \theta g_{\theta\theta}^\dCS\,,
\\
g_{t\phi}^\dCS &=& \frac{5}{8}\zeta M \chi \frac{M^4}{r^4} \left( 1+\frac{12}{7}\frac{M}{r} + \frac{27}{10} \frac{M^2}{r^2} \right) \sin^2 \theta\,,
\end{eqnarray}
to $\mathcal{O}(\zeta,\chi^2)$, with $f(r) \equiv 1-2M/r$. As we can see, the dCS modification deforms the gravitational field of spinning BHs in GR, which is now described by a modified Kerr geometry. 

This solution was extended to $\mathcal{O}(\zeta,\chi^5)$ in Ref.~\cite{Maselli:2017kic} using Hartle-Thorne coordinates. The mapping between Hartle-Thorne $(t,r_{\HT},\theta_{\HT},\phi)$ and Boyer-Lindquist coordinates $(t,r,\theta,\phi)$ to ${\cal{O}}(\chi^2)$ is given by 
\begin{eqnarray}
\label{eqn:TransformationSecondOrder}
r_{\HT} &=& r+\chi ^2  \frac{M^2}{2 r^3}  \Bigg[M^2 (r-M) (2 M+r)\nonumber \\ 
& & +\cos ^2(\theta ) (2 M-r) (3 M+r) \Bigg]\,
\\ 
\theta_{\HT} &=& \theta + \frac{M^2 \sin (2 \theta ) (2 M+r)}{4 r^3}  \chi ^2\,.
\end{eqnarray}
The mapping and the transformed metric metric to ${\cal{O}}(\chi^5, \zeta)$ are given explicitly in~\ref{app:ExplicitMetric}.

%%%%%%%%%%%%%%%%%%%%%%%%%%%%%%%%%%%%%%%%%%%%%
%%%%%%%%%%%%%%%%%%%%%%%%%%%%%%%%%%%%%%%%%%%%%
\section{Theoretical framework}
\label{Framework}
%%%%%%%%%%%%%%%%%%%%%%%%%%%%%%%%%%%%%%%%%%%%%
%%%%%%%%%%%%%%%%%%%%%%%%%%%%%%%%%%%%%%%%%%%%%

Astrophysical bodies cannot be described as simple, structureless test particles in dCS gravity. This is because isolated solutions are not fully determined only by the matter stress-energy tensor, but they may also source a scalar field. No stellar body (including BHs), however, sources a \emph{monopolar} scalar charge in dCS gravity, so they can be roughly approximated as test particles. The motion of small compact objects around a supermassive black hole, an EMRI, can then be approximated as a geodesic of the background spacetime\footnote{When the small object is a BH, the geodesic motion will be corrected by scalar field forces, but we will neglect this effect in this paper and leave it for future work.}~\cite{Sopuerta:2009iy}. Pure geodesic motion in dCS gravity, nonetheless, is not identical to geodesic motion in GR because the background spacetime on which test particles move is not the Kerr spacetime. In this section, we will describe test-particle motion in GR and in dCS backgrounds, as well as a method to characterize chaos in such motion. 

%%%%%%%%%%%%%%%%%%%%%%%%%%%%%%%%%%%%%%%%%%%%%
%%%%%%%%%%%%%%%%%%%%%%%%%%%%%%%%%%%%%%%%%%%%%
\subsection{Geodesic Motion}
\label{GeodesicsMotion}

We can describe geodesic motion through a Hamiltonian function of the form
\begin{equation}
\label{eqn:Hamiltonian}
H=\frac{1}{2\mu}g^{\mu\nu}P_{\mu}P_{\nu}\,,
\end{equation}
where $\mu$ is the rest mass of the orbiting test particle and its corresponding four-momenta is $P_{\mu} \equiv \mu \; u_{\mu}$, with $u_{\mu}$ its four-velocity. The orbital motion is governed by Hamilton's equations 
\begin{equation}
\label{eqn:EOMH}
\dot{q}^{\mu}=\frac{\partial H}{\partial P_{\mu}}, \qquad \dot{P}_{\mu}=-\frac{\partial H}{\partial q^{\mu}}\,,
\end{equation} 
where the overhead dot stands for a derivative with respect to the affine parameter. A Hamiltonian $H \left(q^{\mu},P_{\mu}\right)$ is called \emph{integrable} if one can find a canonical transformation $S\left(q^{\mu},J_{\mu}\right)$ to new variables $(\theta^{\mu}, J_{\mu})$~\cite{arnol2013mathematical} 
\begin{equation}
\label{eqn:Canonical}
\left(q^{\mu},P_{\mu}\right)=\frac{\partial S\left(q^{\nu},J_{\nu}\right)}{\partial q^{\mu}} \longleftrightarrow \left(\theta^{\mu},J_{\mu}\right)=\frac{\partial S\left(q^{\nu}, J_{\nu}\right)}{\partial J_{\mu}}\,,
\end{equation} 
such that in the new coordinates the Hamiltonian depends only on the new momenta $J_{\mu}$. Consequently, the equations of motion in Eq.~(\ref{eqn:EOMH}) in action-angle variables $(\theta^{\mu},J_{\mu})$ are now 
\begin{equation}
\label{eqn:EOMHActionVariables}
\dot{\theta}^{\mu}=\frac{\partial H}{\partial J_{\mu}}= \omega^{\mu}\left(J_{\nu}\right),\qquad \dot{J}_{\mu}=-\frac{\partial H}{\partial \theta^{\mu}}=0\,,
\end{equation} 
which can be immediately integrated to obtain
\begin{eqnarray}
\label{eqn:solutionAV}
\theta^{\mu} & = & \omega^{\mu} \; t + \delta^{\mu}\,, \\
J_{\mu} & = & {\rm{const.}}\,, 
\end{eqnarray}
where $\delta^{\mu}$ and $\omega^{\mu}(J_{\nu})$ are constants. This notion of integrability of Hamiltonian systems is known as \emph{Liouville integrability}~\cite{schuster2006deterministic}: if $n$ linearly independent, integrals of motion exist in a system of $n$ degrees of freedom, then there exists a  coordinate transformation to angle-action variables such that the equations of motion can be put in quadrature form.

Certain symmetries in the spacetimes we typically work with allow us to simplify the evolution of test particles in such a background. The spacetime we study in this paper, presented in Sec.~\ref{IABC}, is stationary and axisymmetric, and thus, the metric tensor is independent of $t$ and $\phi$, and the energy $E=- P_{t}$ and the ($z-$component) orbital angular momentum $L_{z}=P_{\phi}$ are integrals of the motion. The appearance of these two conserved quantities reduces the original four degrees of freedom of the geodesics to only two, represented in the coordinates $\left(r,\theta\right)$. Moreover, since the geodesic equations are autonomous, i.e.~explicitly time-independent, the Hamiltonian of Eq.~(\ref{eqn:Hamiltonian}) is itself an additional constant of motion, whose value is proportional to the rest mass of the test particle $H=-\mu/2$. 

\begin{center}
\begin{figure}[h!]
\begin{center}
{\includegraphics[width=0.5\columnwidth]{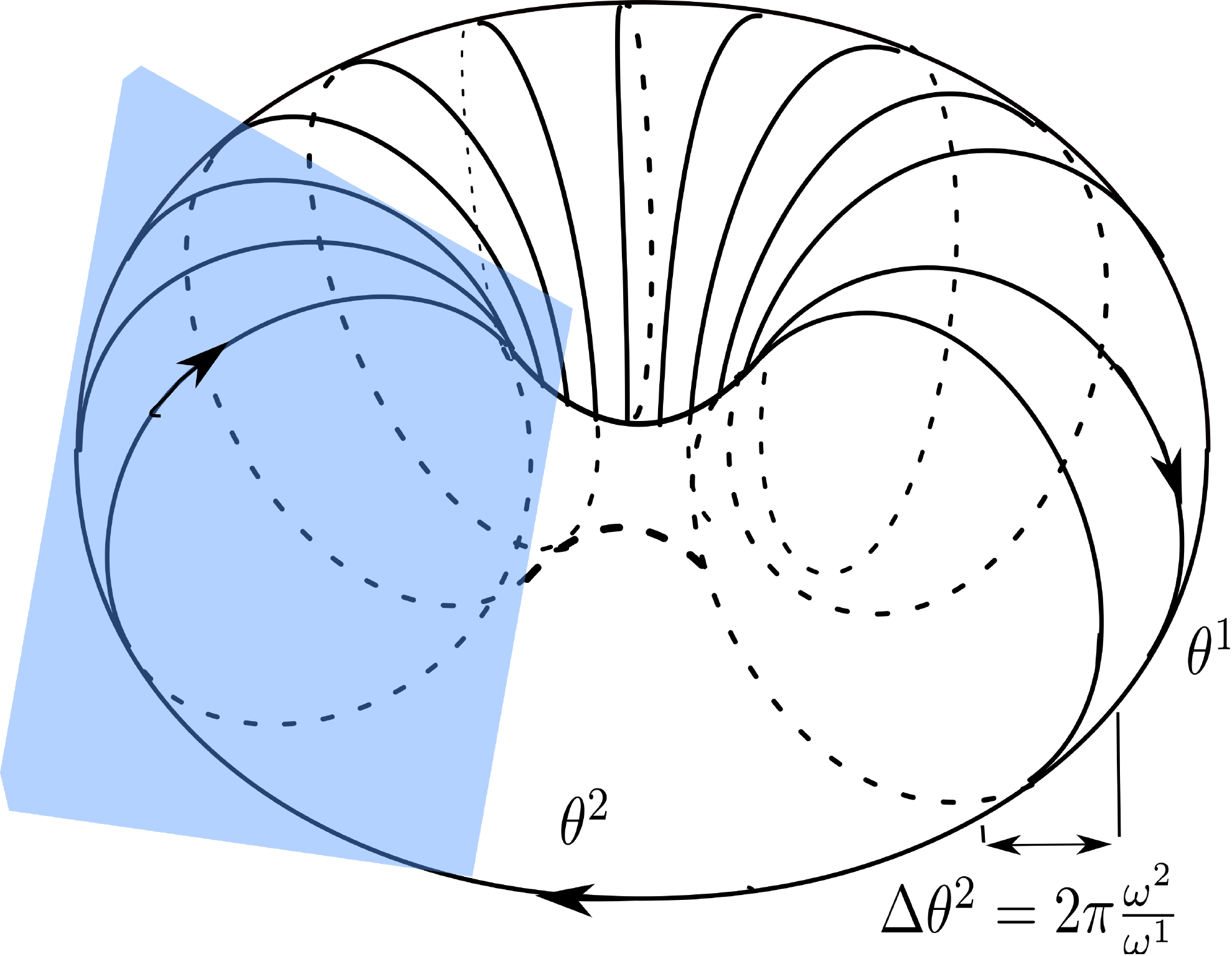}}
\end{center}
\caption{(Color Online) Phase space tori characterized by the 
angle coordinates ($\theta^{1},\theta^{2}$). A phase-orbit is depicted which
wraps around one torus. Closed orbits occur only if the ratio $\frac{\omega^{1}}{\omega^{2}}$ is a rational number.}
\label{Fig:tori1}
\end{figure}
\end{center}

In terms of Liouville integrability, an additional constant of the motion, independent of $(H,E, L_{z})$, is required to guarantee integrability. For example, in the Kerr metric the existence of the Carter constant serves as a fourth, independent constant of the motion~\cite{Carter:1968rr}, and Kerr geodesics are integrable. Therefore, Kerr geodesics exist in a 2-dimensional tori of the 4-dimensional phase space~\cite{LukesGerakopoulos:2010rc}, and bound orbits wrap around this torus with certain characteristic frequencies associated to the angle-action variables~(\ref{eqn:EOMHActionVariables}). A two dimensional slice of these torus is often chosen as a way to map features of the dynamics of the system, as shown in Fig.~\ref{Fig:tori1}, where the two dimensional slice is shown in blue.

\begin{center}
\begin{figure}[h!]
\begin{center}
\includegraphics[width=0.8\columnwidth]{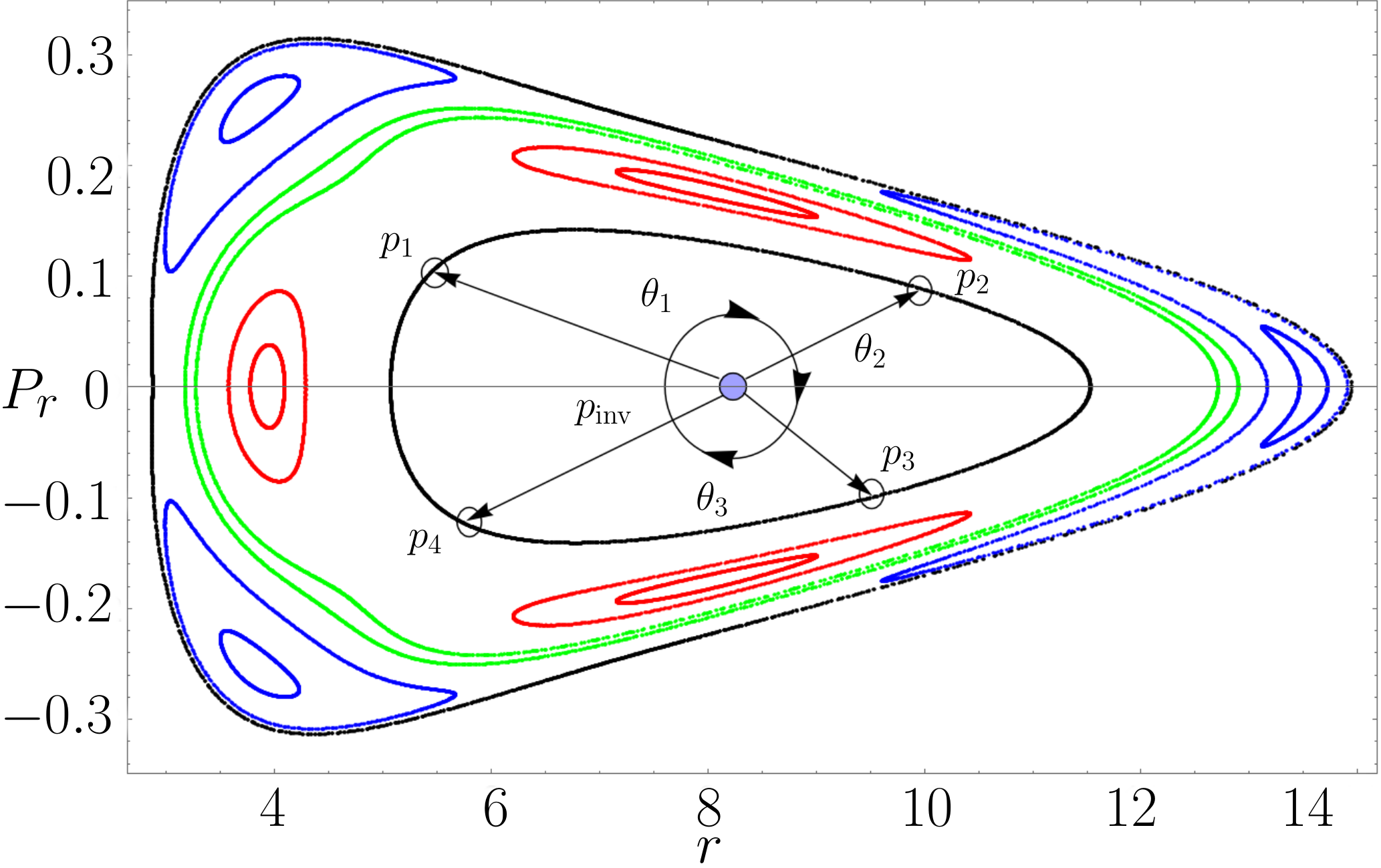}
\end{center}
\caption{(Color Online) Poincar\'e surfaces of section with the chaotic features of interest. The inner part of the figure shows the procedure to compute the rotation number. The purple circle indicates the position of the invariant point, and each of the arrows point towards the position of a crossing in the Poincar\'e map, denoted with circles. }
\label{Fig:rotnumber}
\end{figure}
\end{center}

A \emph{Poincar\'e surface of section} is defined by the successive intersections of the orbit with a chosen two dimensional slice of the torus. Each time the geodesic pierces this slice, a single point is generated on the slice~\cite{lichtenberg1992regular}. The complete Poincar\'e surface of section is therefore produced by a sufficient number of successive piercings, as shown in Fig.~\ref{Fig:rotnumber}. 

For an integrable system, all initial conditions generate either periodic, or quasi-periodic trajectories. The difference between one initial condition and the next manifests as a monotonic change in the ratio of the polar to the radial frequency of the motion, $\omega^{\theta}/\omega^{r}$, provided that the non-degeneracy condition holds, e.g., a nonzero determinant $\left|\partial\omega^{\mu}/\partial J_{\mu}\right|$ \cite{riissmann1989non}. Periodic orbits happen when this ratio is a rational number $n/m$, with $(n,m) \in {\mathbb{N}}$, and the phase-orbit repeats itself after $m$ windings. On the other hand, quasi-periodic orbits occur when the ratio of these frequencies, $\omega^{\theta}/\omega^{r}$, is an irrational number and the phase-orbit is densely covered~\cite{LukesGerakopoulos:2010rc}.

Let us now consider a Hamiltonian that is a deformation of an integrable Hamiltonian
\begin{equation}
H = H_{\Inte} + \delta H\,.
\end{equation}
Typically, such a deformed Hamiltonian can be treated within Hamilton-Jacobi theory as described above, but only upon orbit-averaging~\cite{schuster2006deterministic} as the treatment fails for periodic orbits, which by definition contain resonances of the motion. If the deformation is non-integrable, then the full motion is non-integrable, and regular or completely irregular motion can result depending on the initial conditions. 

For a non-integrable deformation, the behavior of the phase-orbits depends on its magnitude. If the deformation is large, then the phase space portrait changes significantly, destroying the phase space portrait, which carries imprints in the Poincar\'e surfaces of section, and signaling the presence of strong chaotic behavior. 

Conversely, if the deformation is small, then the Kolmogorov-Arnold-Moser (KAM) theorem~\cite{schuster2006deterministic} implies that some of the invariant tori are deformed and survive, while others are destroyed. In other words,  the corresponding Poincar\'e surfaces of section of the integrable and of the deformed systems look very similar to each other. For quasi-periodic orbits, the resulting deformed phase-orbits are called \emph{KAM curves}. The green curves shown in Fig.~\ref{Fig:rotnumber} are deformed tori that correspond to quasiperiodic orbits.

The behavior of periodic orbits is more complicated and can be understood through the Poincar\'e-Birkhoff theorem~\cite{lichtenberg1992regular}. The tori formed by periodic orbits, the so-called \emph{resonant torii}, will dissolve after the perturbation and can even break into a finite number of periodic points, half of them elliptic (stable) and half of them hyperbolic (unstable), distributed in an alternating pattern. Surrounding the elliptic points, a set of small KAM curves appear, called \emph{Bifkhoff islands} and depicted schematically through the red and blue nested structures in Fig.~\ref{Fig:rotnumber}. These elliptic points are the source of asymptotic manifolds which are the underlying structure of chaos. This chain can cover the phase space, without necessarily existing in a preferred location, which is why they are interweaved with regular regions~\cite{schuster2006deterministic}. 

As we discussed in Sec.~\ref{IABC}, the background spacetime in dCS gravity can be considered as perturbations of spacetimes in GR, i.e., the Schwarzschild metric and the Kerr metric in the resummed case, both of which lead to integrable geodesics in the $\zeta \to 0$ limit. The Hamiltonian of Eq.~(\ref{eqn:Hamiltonian}) can then be written as
\begin{equation}
\label{eqn:PertrubedHamiltonian}
H=\frac{1}{2\mu}g_{\GR}^{\mu\nu}P_{\mu}P_{\nu}+\frac{1}{2\mu} \; \epsilon \; g_{\dCS}^{\mu\nu}P_{\mu}P_{\nu}+\mathcal{O}\left(\epsilon^{2}\right), 
\end{equation} 
where $\epsilon$ is a book-keeping parameter that labels the order of the dCS perturbation. Such a deformed Hamiltonian can be treated within Hamilton-Jacobi theory, such as described above, upon orbit-averaging~\cite{Yagi:2012ya}, but again, this treatment is ill suited for periodic orbits. Therefore, to fully understand the phase space portrait generated by geodesics in dCS gravity, one must resolve and characterize the phase space structure\footnote{This type of techniques have been already used extensively in the literature to spot chaotic behavior~\cite{contopoulos2013order, Dubeibe:2007xka, Gair:2007kr, Brink:2008xy,LukesGerakopoulos:2012pq, LukesGerakopoulos:2010rc, Zelenka:2017aqn,Lukes-Gerakopoulos:2017jub}. }. If the perturbation to the Hamiltonian introduced by dCS gravity is non-integrable, we then expect either deformed invariant tori, or the appearance of Birkhoff islands in the phase space portrait, in both cases governed by the magnitude of the coupling parameter $\zeta$.

%%%%%%%%%%%%%%%%%%%%%%%%%%%%%%%%%%%%%%%%%%%%%
%%%%%%%%%%%%%%%%%%%%%%%%%%%%%%%%%%%%%%%%%%%%%
\subsection{The Rotation Number}

Despite the fact that the Poincar\'e surface of sections display all the features we are interested in, e.g.~Birkhoff islands or deformed tori, these regions can be very small in phase space, and thus, very hard to spot. Fortunately, there is a very powerful tool, the \emph{rotation number}~\cite{contopoulos2013order,LukesGerakopoulos:2010rc}, that allows us to study quantitatively the characteristics of chaos.

The rotation number corresponding to a Poincar\'e surface of section can be calculated by first identifying the central invariant point of the section~\cite{LukesGerakopoulos:2010rc}. This is the fixed point corresponding to the periodic orbit which crosses the two-dimensional slice defining the Poincar\'e surface, which for us will be the equatorial plane $\theta=\pi/2$, at only one point with $P_{r}=0$ (see the purple circle shown in Fig.~\ref{Fig:rotnumber}). With the invariant point identified, the rotation number can be computed as follows. Consider two vectors in phase space joining the invariant point, $p_{\inv}=\left(r_{\inv},P_{r_{\inv}}\right)$, to two successive piercings of the two dimensional surface slice, labeled as $p_j=\left(r_{j},P_{r_{j}}\right)$ and $p_{j+1}=\left(r_{j+1},P_{r_{j+1}}\right)$. These vectors defined from the origin $(r=0,P_{r}=0)$ are
\begin{equation}
 \vec{A}_{j}=\vec{p}_{j}-\vec{p}_{\mathrm{inv}},
 \qquad
  \vec{A}_{j+1}=\vec{p}_{j+1}-\vec{p}_{\mathrm{inv}}.
\end{equation}
From these vectors we now calculate the clockwise angle subtended by them, i.e., 
\begin{equation}
 \theta_{j}=\measuredangle (\vec{A}_{j+1}, \vec{A}_{j}),
\end{equation}
as shown in Fig.~\ref{Fig:rotnumber} for four points $p_{i}$ that belong to the same Poincar\'e surface section. This angle is computed for each consecutive pair of piercings. Summing up all these angles $\theta_{j}$ and dividing by $2\pi N$, with $N$ the number of piercings in the corresponding section, the rotation number is then 
\begin{equation}
 \nu_{\theta}=\displaystyle{\lim_{N \to \infty}} \; \frac{1}{2\pi 
N}\sum^{N}_{j=1} \; \theta_{j}.
\label{rotnum}
\end{equation}

\begin{center}
\begin{figure}[h!]
\begin{center}
\includegraphics[width=0.7\columnwidth]{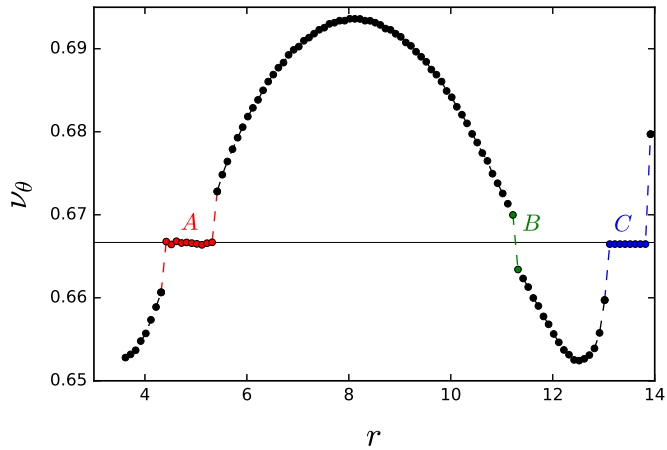}
\end{center}
\caption{(Color Online) The rotation curve of the example shown in Fig.~\ref{Fig:rotnumber}. The different type of behaviors that can be seen using the rotation number are marked as $A$ (plateau near a stable periodic orbit), $B$ (an inflection point at an unstable periodic orbit) and $C$ (plateau near a stable periodic orbit).}
\label{Fig:rotcurve}
\end{figure}
\end{center}

The rotation number characterizes the frequency structure of the phase space for each trajectory and measures the average fraction of a circle by which successive crossings advance~\cite{LukesGerakopoulos:2010rc}. The  \emph{rotation curve} of the system is obtained by evaluating the rotation number as a function of the location of the Poincar\'e surface of section in phase space. In our case, we choose to define the location of the surface by the minimum value of the radial coordinate sampled by that surface. The rotation curve associated with the example depicted in Fig.~\ref{Fig:rotnumber} is shown in Fig.~\ref{Fig:rotcurve}, where the colors have been chosen to match the regions of interest. Birkhoff islands corresponding to a plateau in the rotation curve are presented in red and blue, and an abrupt change of the rotation number near an unstable point, corresponding to the regions where the tori structure is deformed but not broken, is shown in green. For regular regions, we can see that the rotation curve looks smooth. 

Abrupt changes in the rotation curve signal the presence of chaotic orbits. In the region where regular orbits exist, the rotation curve is a monotonically increasing function of the spatial coordinate. Inside Birkhoff islands, however, each member of the chain has a fixed rational value of the frequency ratio, forcing the rotation number to remain approximately constant, up to numerical accuracy,  and one finds a \emph{plain zone} or \emph{plateau}. The variation of the rotation number between nearby chaotic orbits is completely irregular and not defined unambiguously, breaking also the monotonicity of the curve~\cite{contopoulos2013order}.

The usefulness of the rotation number goes beyond its ability to identify the dynamics of a system. For example, a plateau in the rotation number signals the presence of a constant ratio of the orbital frequencies, which then translates into a constant pattern of frequencies in the emitted gravitational waves. An observation of such a constant pattern would constitute a clear signal of the presence of chaos, and thus, a novel test of GR and the Kerr hypothesis~\cite{lukes2013mind,Gair:2007kr,LukesGerakopoulos:2010rc,Lukes-Gerakopoulos:2017jub}. 

%%%%%%%%%%%%%%%%%%%%%%%%%%%%%%%%%%%%%
%%%%%%%%%%%%%%%%%%%%%%%%%%%%%%%%%%%%%
\section{Geodesics in the slowly-rotating Kerr metric}
\label{Geodesics}

%%%%%%%%%%%%%%%%%%%%%%%%%%%%%%%%%%%%%
%%%%%%%%%%%%%%%%%%%%%%%%%%%%%%%%%%%%%

As a warm-up to the dCS problem, let us first consider geodesics in the Kerr background and in its slow-rotation expansion. The latter can be obtained by expanding the Kerr metric in $\chi \ll 1$ to any order one wishes. At zeroth order in rotation, the slow-rotation expansion of the Kerr metric reduces exactly to the Schwarzschild metric, which is integrable. At next order in rotation, the slow-rotation expansion can be thought of as a deformation of Schwarzschild, which can (and in fact does) generate chaotic orbits. Of course, as one keeps higher and higher order terms in the slow-rotation expansion, one expects the resulting metric to become closer and closer to the Kerr spacetime, with therefore any signs of chaos shrinking with the order retained. This is the behavior we expect, find and explore in this section.    

We study geodesics of the slow-rotation expansion to ${\cal{O}}(\chi^{n})$ treating the metric as exact, i.e.~once the metric is expanded, it is treated as exact and the geodesic equations are solved numerically without re-expanding them in slow-rotation. We make this choice based on numerical explorations where we compute the rotation curve using a slowly-rotating Kerr metric treated as (i) exact and (ii) approximate, i.e.~re-expanding the geodesic equations in small rotation prior to solving them numerically. For all cases studied, we found that the relative error of (i) is several orders of magnitude smaller than that of (ii), relative to the rotation curve computed with the full Kerr metric. When dealing with case (i), the geodesic equations contain a plethora of higher-order-in-spin terms, which method (ii) sets to zero. For the rotation curve, these higher order in spin terms introduced in case (i) happen to lead to a smaller error. This may not be true for other observables, as a generic statement cannot be made. Moreover, we find that treating the slowly-rotating Kerr metric as exact does not introduce any pathologies.

In what follows, we focus mostly on resonant orbits in a slowly-rotating Kerr metric expanded to ${\cal{O}}(\chi^{2})$. Even though we will show results for mainly two representative examples of the background (with $\chi=0.1$ and $\chi=0.2$ and one particular resonance), the features we find are generic and based on an extensive numerical study in a very large region of parameter space. One may worry that a choice of $\chi = 0.2$ is inappropriate for a slow-rotation expansion at ${\cal{O}}(\chi^{2})$, but (i) this choice is actually conservative relative to other studies that have used slowly-rotating metrics in the past~\cite{Canizares:2012is,stein2014parametrizing, vincent2013testing, Maselli:2017kic, ayzenberg2017black}, and (ii) a smaller choice of $\chi$ does not eliminate the features we will discuss here, but instead just makes them more difficult to resolve numerically. 

Let us begin the discussion of geodesics by studying the regime in phase space where orbits exist. From the normalization relation $u^{\alpha}u_{\alpha}=-1$, the geodesic motion in the reduced system of two degrees of freedom described in the previous section is characterized by the two-dimensional effective potential 
\begin{equation}
\label{eqn:EffectivePotential}
V_{eff}=\frac{1}{2}\left(\frac{g_{\phi\phi}E^{2}+2g_{t\phi}EL_{z}+L^{2}g_{tt}}{g_{tt}g_{\phi\phi}-g_{t\phi}^{2}}+\mu^2\right).
\end{equation} 
The effective potential is characterized by parameters that depend on the spacetime, such as the spin and mass (and also $\zeta$ in dCS gravity), and also on the particle's conserved energy and the $z$-component of the conserved angular momentum vector. When $P_{r}=0=P_{\theta}$, the roots of the effective potential $V_{\eff}\left(r,\theta,E,L_{z}\right)=0$ define the so-called \emph{curves of zero velocity} (CZV)~\cite{Gair:2007kr}.

\begin{figure*}[hpt]
\begin{tabular}{ll}
\includegraphics[width=0.5\columnwidth{}]{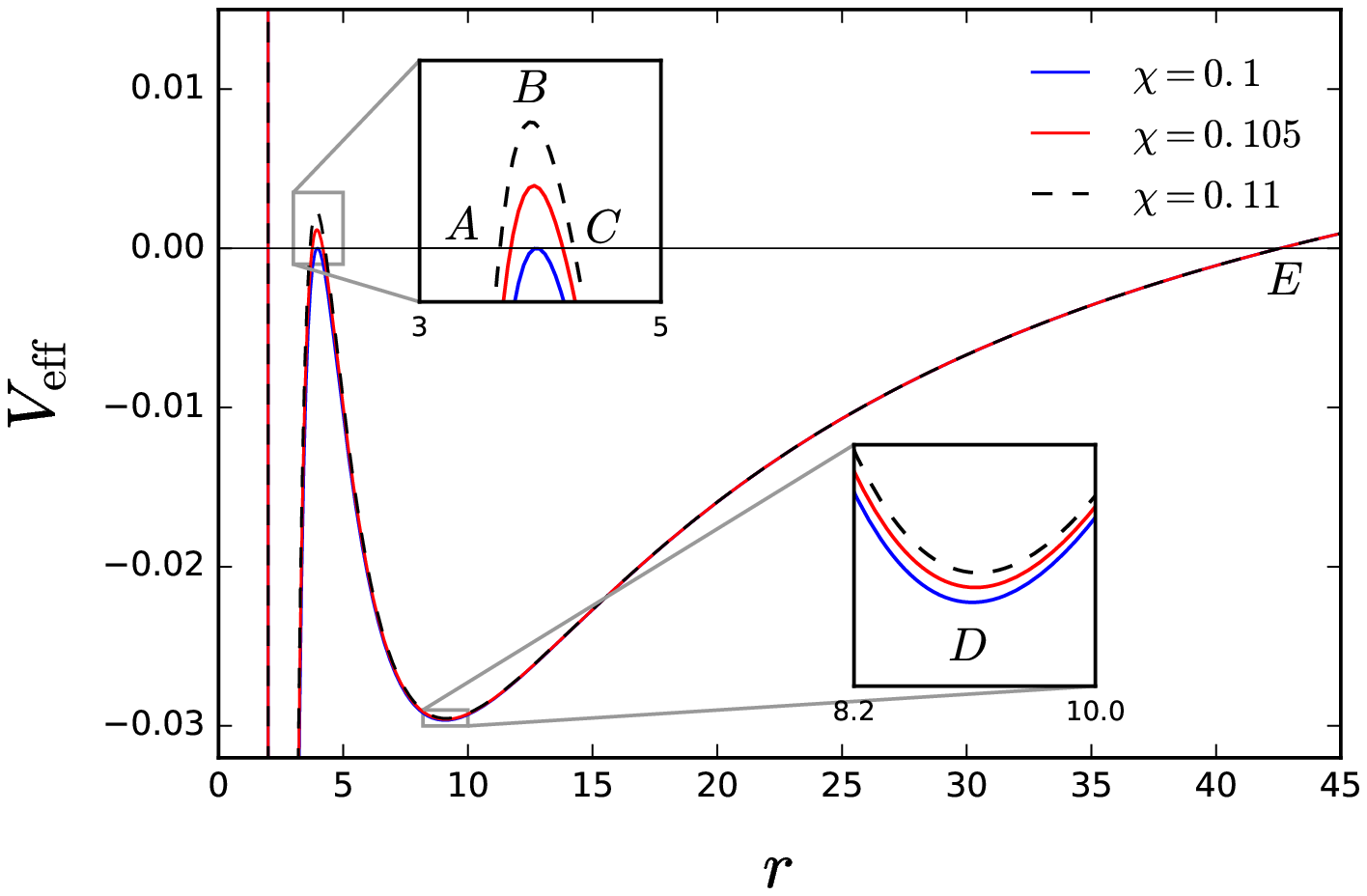}
%&
\includegraphics[width=0.5\columnwidth{}]{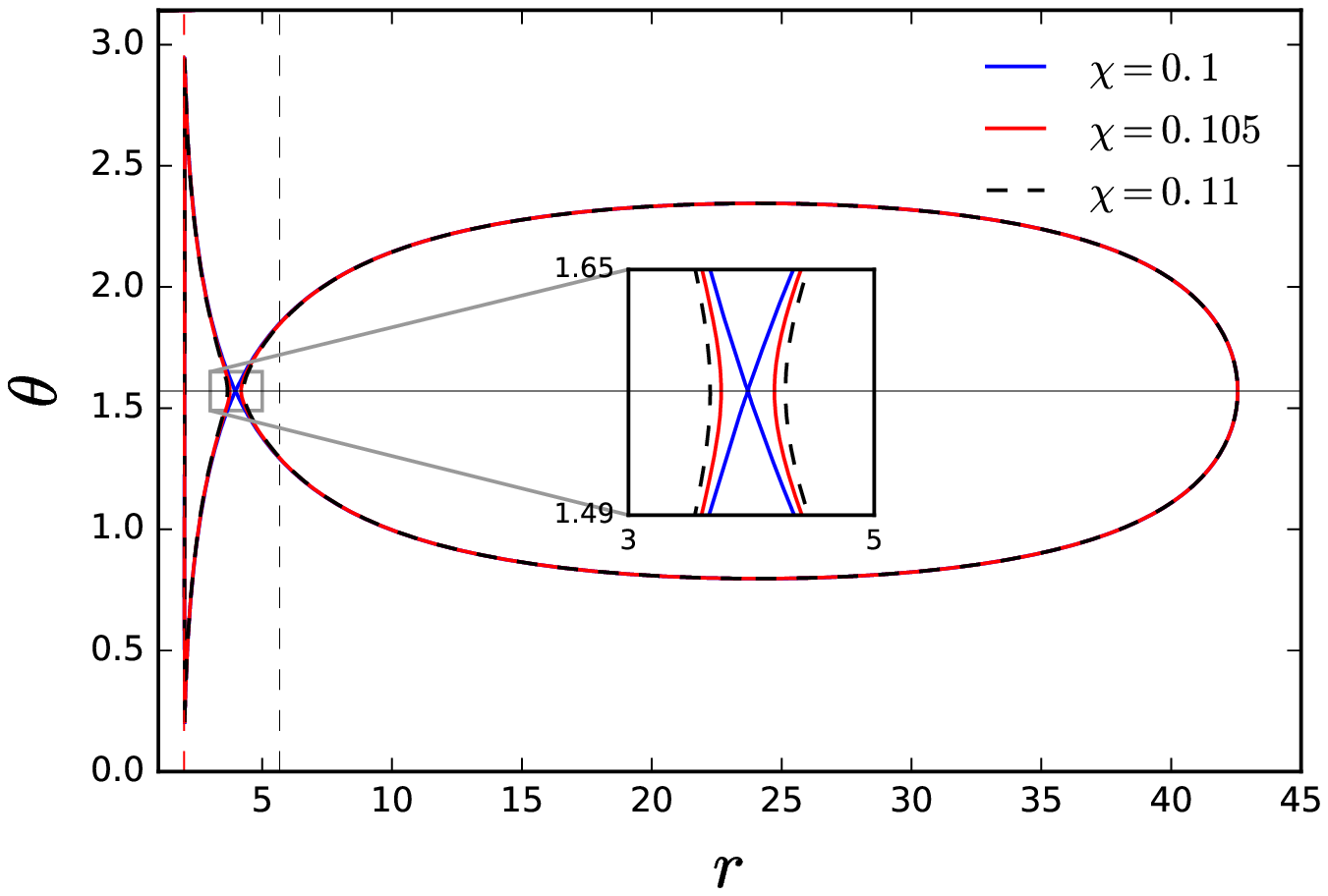} 
\end{tabular}
\caption{(Color Online) (Left)
Two-dimensional potential $V_{\eff}(r,\theta)$ along $\theta=\pi/2$ and (right) CZVs using the slowly-rotating Kerr metric expanded to ${\cal{O}}(\chi^{2})$ for three different values of the spin, at a fixed energy of $E=0.98$ and angular momentum of $L_{z}=3.74265M$. The vertical dashed line in the right panel represents the location of the innermost stable circular orbit, or ISCO, when $\chi=0.1$. The roots of $V_{\eff}$ provide boundaries for different types of orbits.} 
\label{fig:Potentials}%
\end{figure*}

The left and right panels of Fig.~\ref{fig:Potentials} show $V_{\eff}$ and the CZVs, respectively, using the Kerr metric expanded to ${\cal{O}}(\chi^{2})$ as a function of radius and for different choices of the spin parameter. The roots of $V_{\eff}$ provide boundaries for different types of orbits~\cite{lukes2013mind}: plunging orbits (from the horizon to the point labeled $A$ in the figure), bounded non-plunging orbits (between the points labeled $C$ and $E$ in the figure), and periodic orbits (the points labeled $A$, $C$ and $E$ in the figure). We call the latter periodic because the points $B$ and $D$ are local extrema of $V_{\eff}$, a maximum (representing an unstable periodic orbit) and a minimum (representing a stable periodic orbit), respectively. The classification of orbits can also be inferred from the right panel of Fig.~\ref{fig:Potentials}, where two disconnected regions appear. Particles inside the left one are bounded by the horizon, i.e., plunging orbits, and particles inside the right one are bounded but in non-plunging orbits. 

We integrate the equations of motion in Eqs.~(\ref{eqn:EOMH}), for the variables $\left(\dot{r},\dot{\theta},\dot{P}_{r},\dot{P}_{\theta}\right)$, numerically using an explicit Runge$-$Kutta method due to Dormand and Prince~\cite{dormand1980family} (DOPRI). Although Runge$-$Kutta methods are dissipative~\cite{bodenheimer2006numerical}, the DOPRI method can be used here without compromising the conclusions extracted to the required precision and evolution length we consider. In fact, for the orbits under consideration, the conserved quantities remain constant under numerical evolution, ensuring that the particle does not wander in phase space due to numerical error. This algorithm has been already used for studies of chaotic behavior in~\cite{Apostolatos:2009vu,Brink:2015roa}. 

With the background spacetime, the equations of motion and an integration scheme defined, we now proceed to study the dynamics of test particles in detail and compute Poincar\'e  surfaces of section and rotation curves. We will split our analysis into a study of bounded orbits and one of unbounded orbits.

%---------------------------------------------------------------------------------------------------------------------------------------
\subsection{Bound Orbits}
\label{sec:bounded-orbits}

For bounded motion, the test particle is confined to a particular region of the effective potential. Within this region, there is a local minimum, depicted in detail in the right-most embedded diagrams of the left panel of Fig.~\ref{fig:Potentials}, implying the existence of stable periodic orbits. In the same sense as in Newtonian gravity, the angular momentum of the test particle is responsible for a centrifugal barrier that prevents the particle from falling into the central object. Here, since the SMBH background is spinning, there is an extra contribution to the centrifugal barrier due to the BH spin.
 
\begin{figure*}[hpt]
\begin{tabular}{ll}
\includegraphics[width=0.5\columnwidth{}]{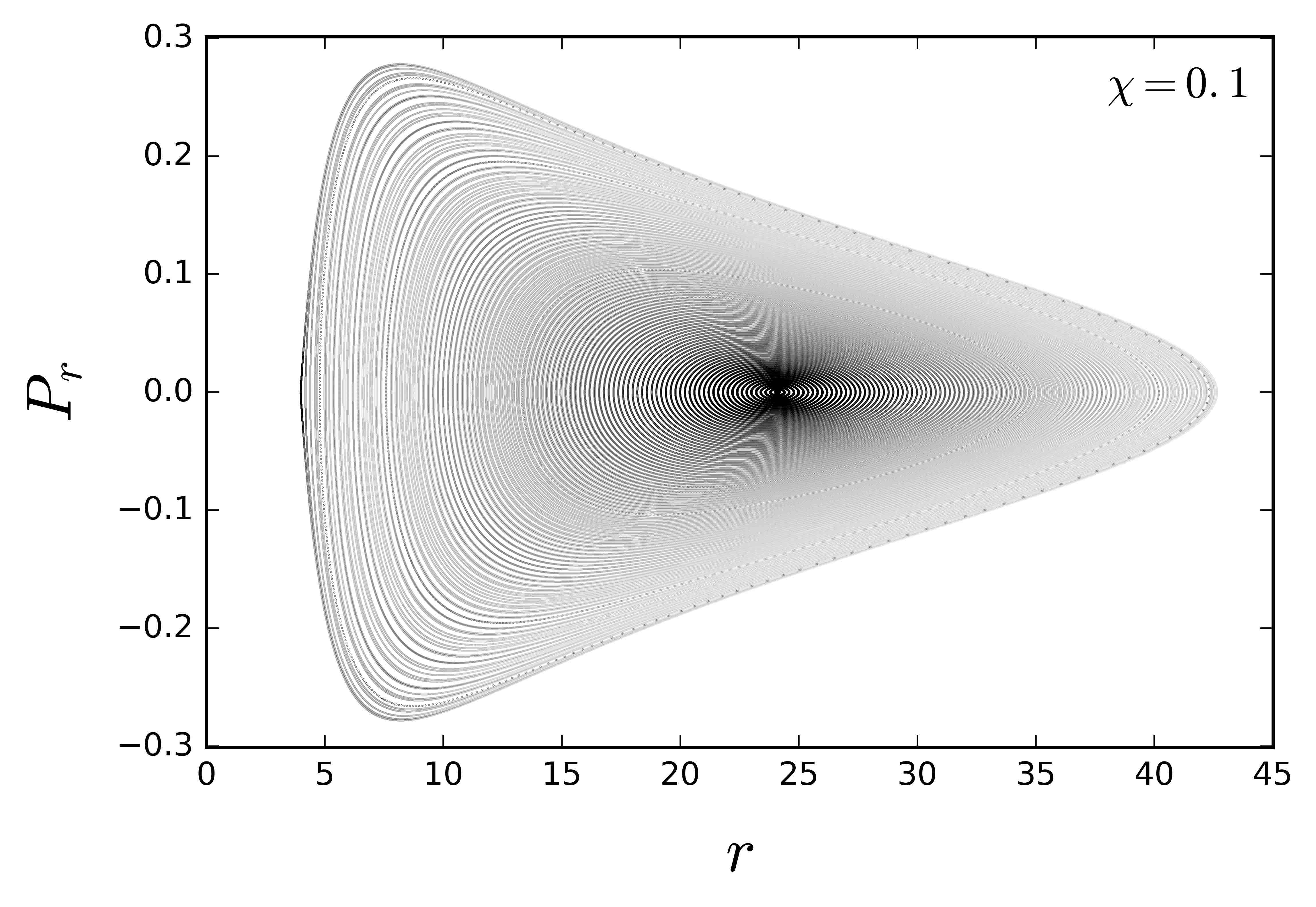}
&
\includegraphics[width=0.5\columnwidth{}]{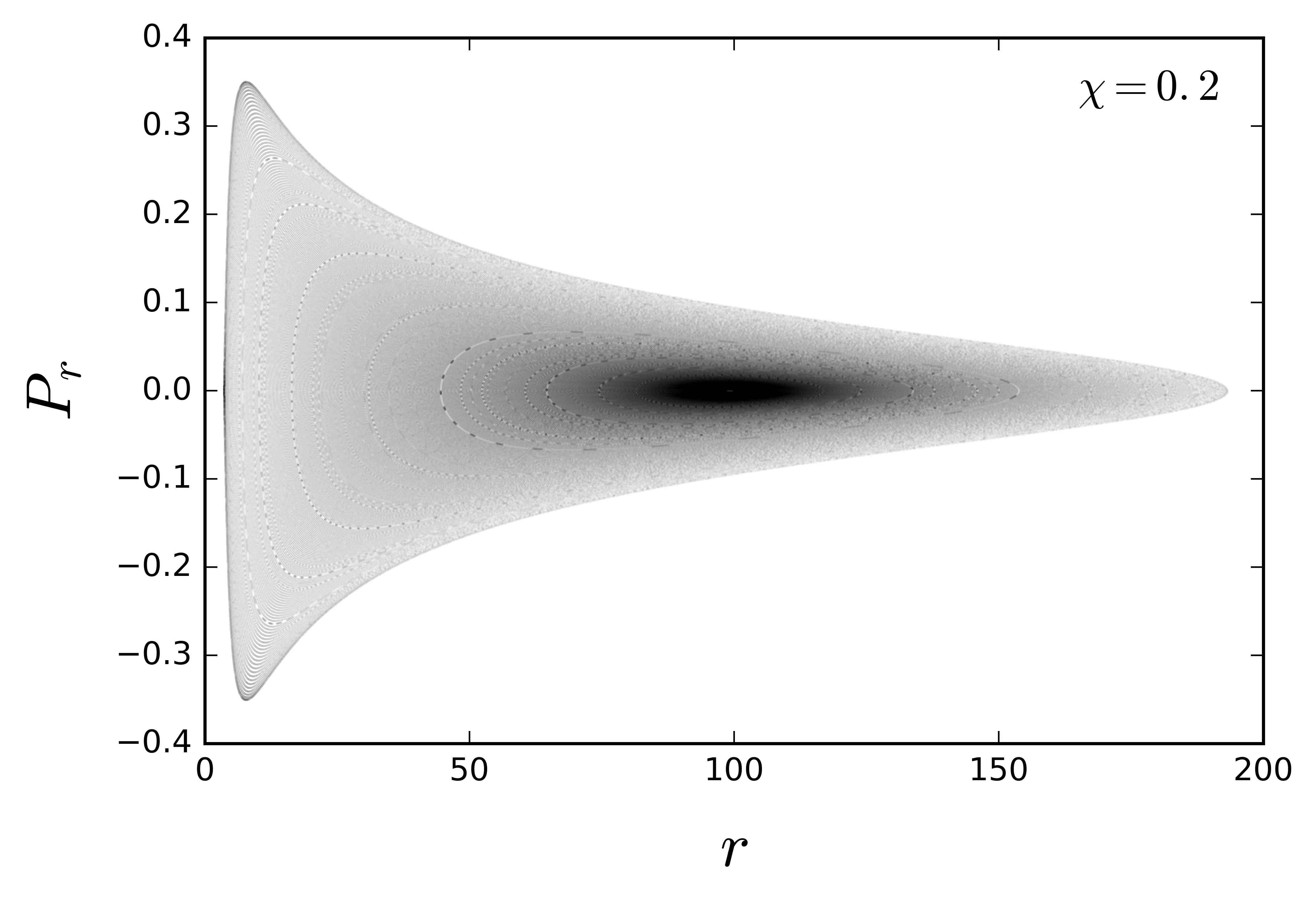}
\end{tabular}
\caption{Poincar\'e surfaces of section for a SMBH with $\chi=0.1$, $E=0.98$ and angular momentum $L_{z}=3.74265M$ (left), and $\chi=0.2$, $E=0.995$ and angular momentum $L_{z}=3.75365M$ (right), using the slow-rotation expansion of the Kerr metric to ${\cal{O}}(\chi^{2})$.} 
\label{fig:Sectionsa01a02}%
\end{figure*}

Bearing in mind the discussion in Sec.~\ref{GeodesicsMotion} about perturbed systems, we expect geodesics of a slowly-rotating Kerr metric to be chaotic. Figure~\ref{fig:Sectionsa01a02} shows Poincar\'e  surfaces of section for two different sets of parameters that produce bounded orbits, using the slowly-rotating metric to ${\cal{O}}(\chi^{2})$. The surfaces of section look regular, seemingly without any signatures of chaos, but this can be deceiving because features of chaos may be small and hard to resolve on this scale.  

\begin{figure*}[hpt]
\begin{center}
\begin{tabular}{ll}
\includegraphics[width=0.5\columnwidth{}]{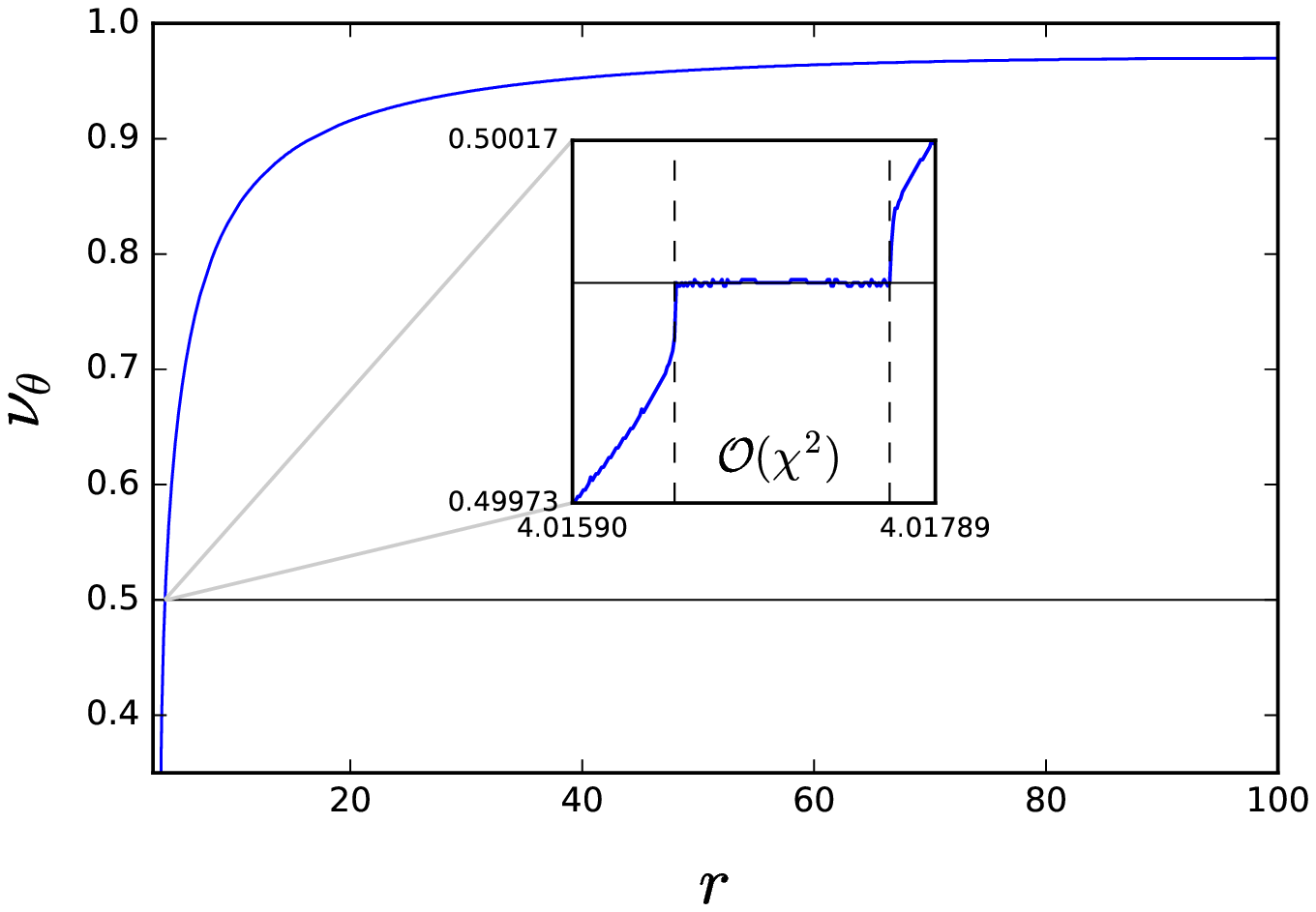}
&
\includegraphics[width=0.5\columnwidth{}]{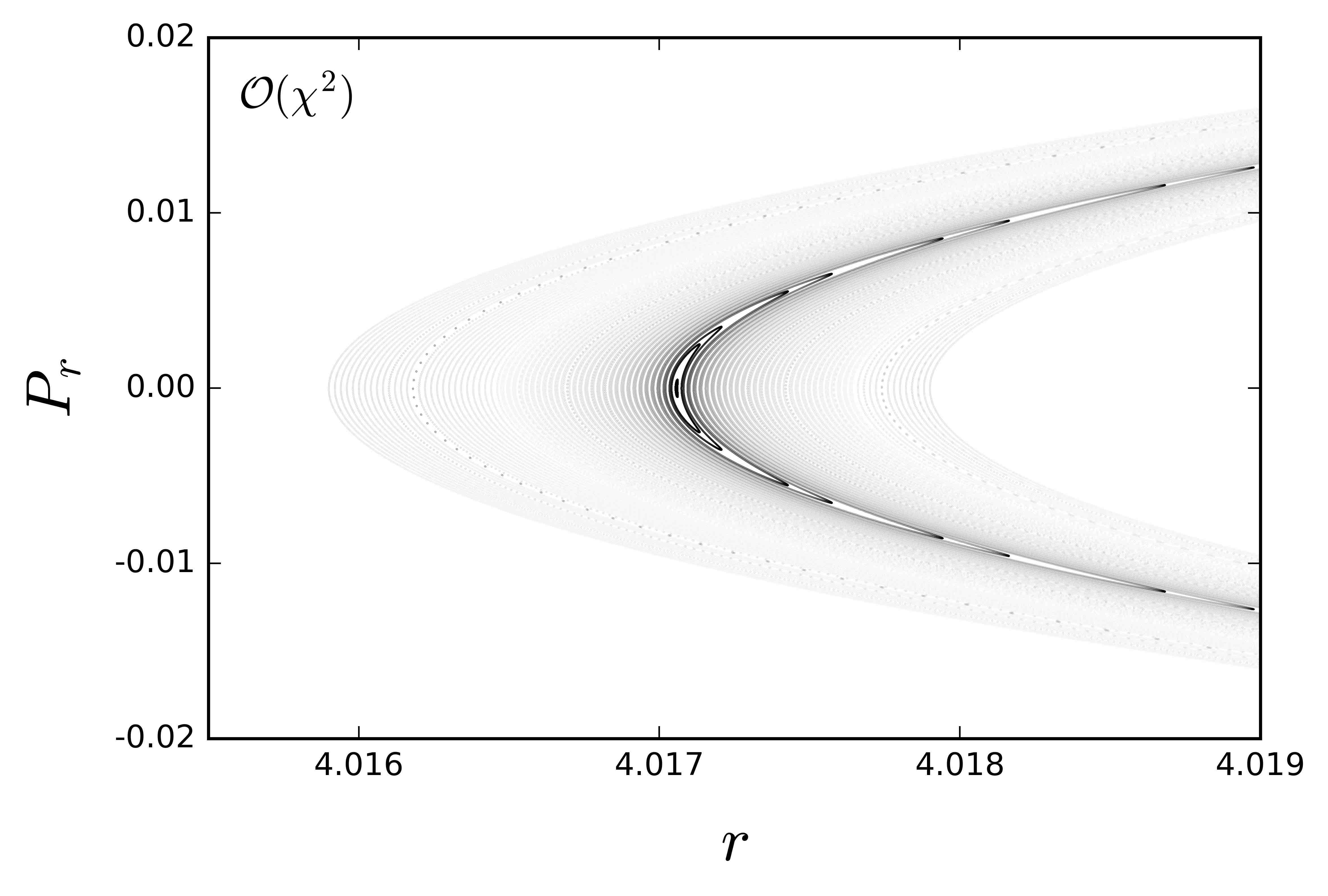}
\end{tabular}
\end{center}
\caption{(Color Online) Rotation curve (left) and surfaces of section (right) for geodesics with $E=0.995$ and angular momentum $L_{z}=3.75365M$ around a SMBH with $\chi=0.2$, using a metric expanded to ${\cal{O}}(\chi^{2})$. Embedded in the left panel is a zoom of the rotation curve around a regime that presents chaotic features. The horizontal line is drawn at $\nu_{\theta}=0.5$. 
} 
\label{fig:RotCurves}%
\end{figure*}

The rotation curve, however, can signal the presence of chaotic behavior, even when this cannot be resolved with the naked eye from the surfaces of section. The left panel of Fig.~\ref{fig:RotCurves} shows the rotation curve for the same sets of orbits as those in the right-panel of Fig.~\ref{fig:Sectionsa01a02}. Observe that there is a plateau in the rotation number around $r \approx 4.016 M$. In practice, such behavior is found by studying the derivative of the rotation number with respect to radius. This plateau is a tell-tale sign of chaos, which we can use to refine our search to a subregion of the surfaces of section. Doing so, we find Birkhoff chains of islands in the surfaces of section, as shown in the right panel of Fig.~\ref{fig:RotCurves}.

\begin{figure*}[hpt]
\begin{tabular}{ll}
\includegraphics[width=0.5\columnwidth{}]{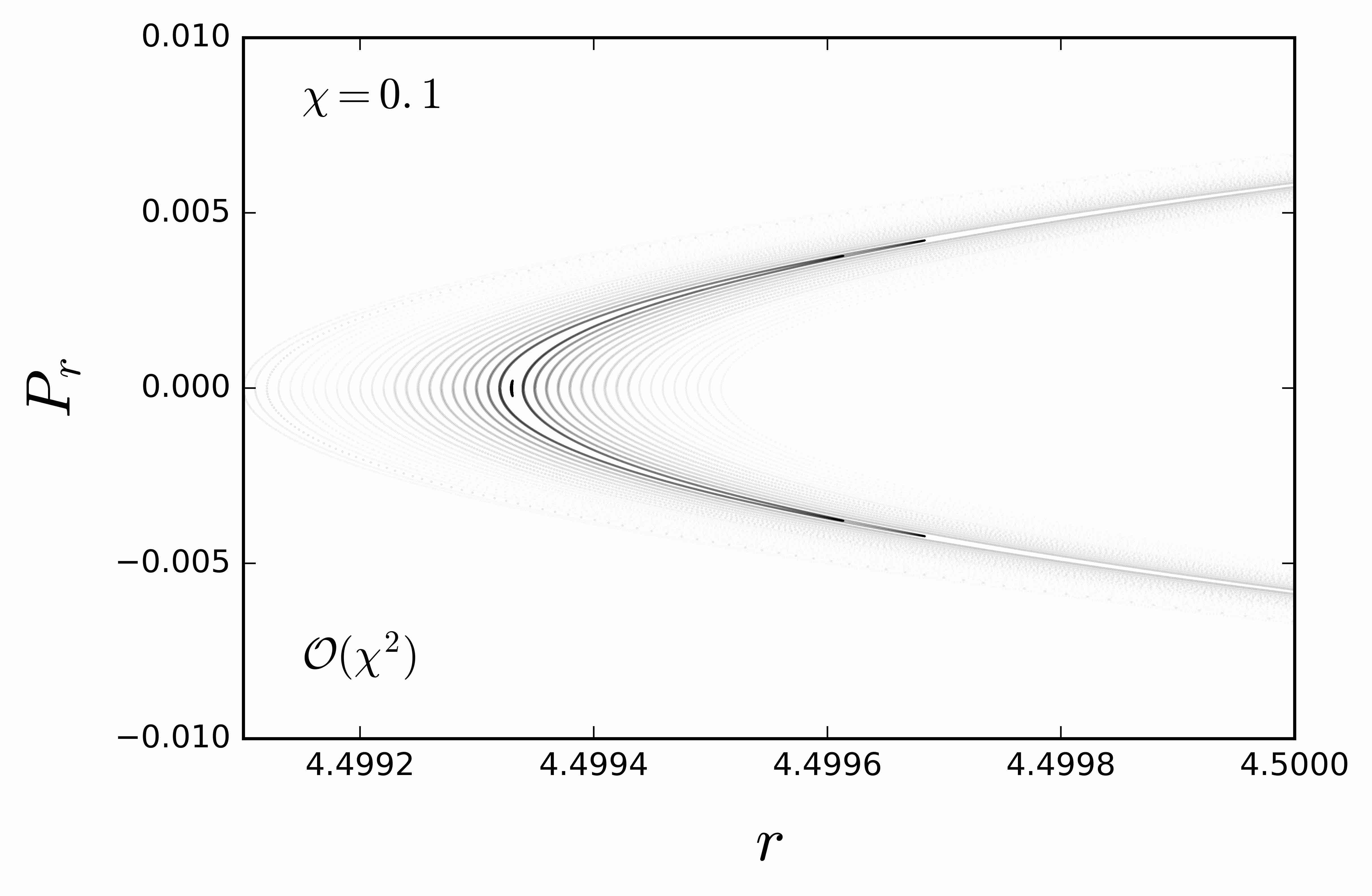} 
&
\includegraphics[width=0.47\columnwidth{}]{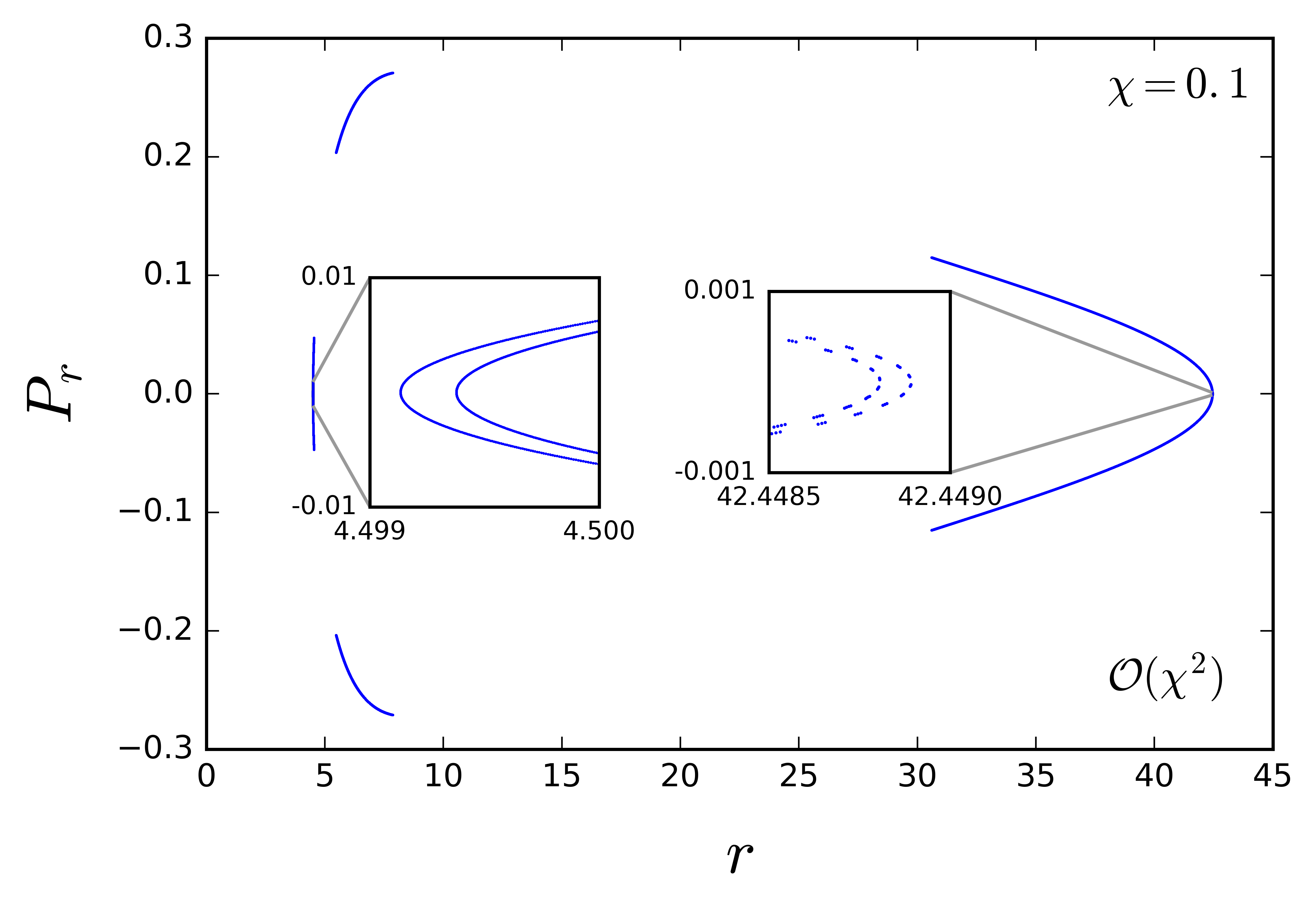}
\end{tabular}
\caption{(Color Online) Birkhoff islands associated with the $2/4$-resonance using the slow-rotation expansion of the Kerr metric to ${\cal{O}}(\chi^{2})$ with $\chi=0.1$, energy $E=0.98$ and angular momentum $L_{z}=3.74265M$. The left panel shows many chain of islands around a stable point. The right panel shows how the structure of the broken tori for one initial condition and the appearance of the islands of stability, in this case four are formed.} 
\label{fig:24Resonancea01}
\end{figure*}

The emergence of Birkhoff islands is generic and not dependent on the particular parameters of the geodesic or of the SMBH. The left-panel of Fig.~\ref{fig:24Resonancea01} is a zoom of the left-panel of Fig.~\ref{fig:Sectionsa01a02} to a regime where the rotation number suggests the presence of chaos. As expected, we find Birkhoff islands once more, although the decrease in the spin parameter has led to fewer islands than in the larger spin case. Zooming back out, we now see that the tori structure is broken and four stable points appear, which are associated with the rational frequency $2/4$-resonance, as predicted by the Poincar\'e$-$Birkhoff theorem and shown in the right panel of Fig.~\ref{fig:24Resonancea01}. 

\begin{figure*}[hpt]
\begin{center}
\begin{tabular}{ll}
\includegraphics[width=0.5\columnwidth{}]{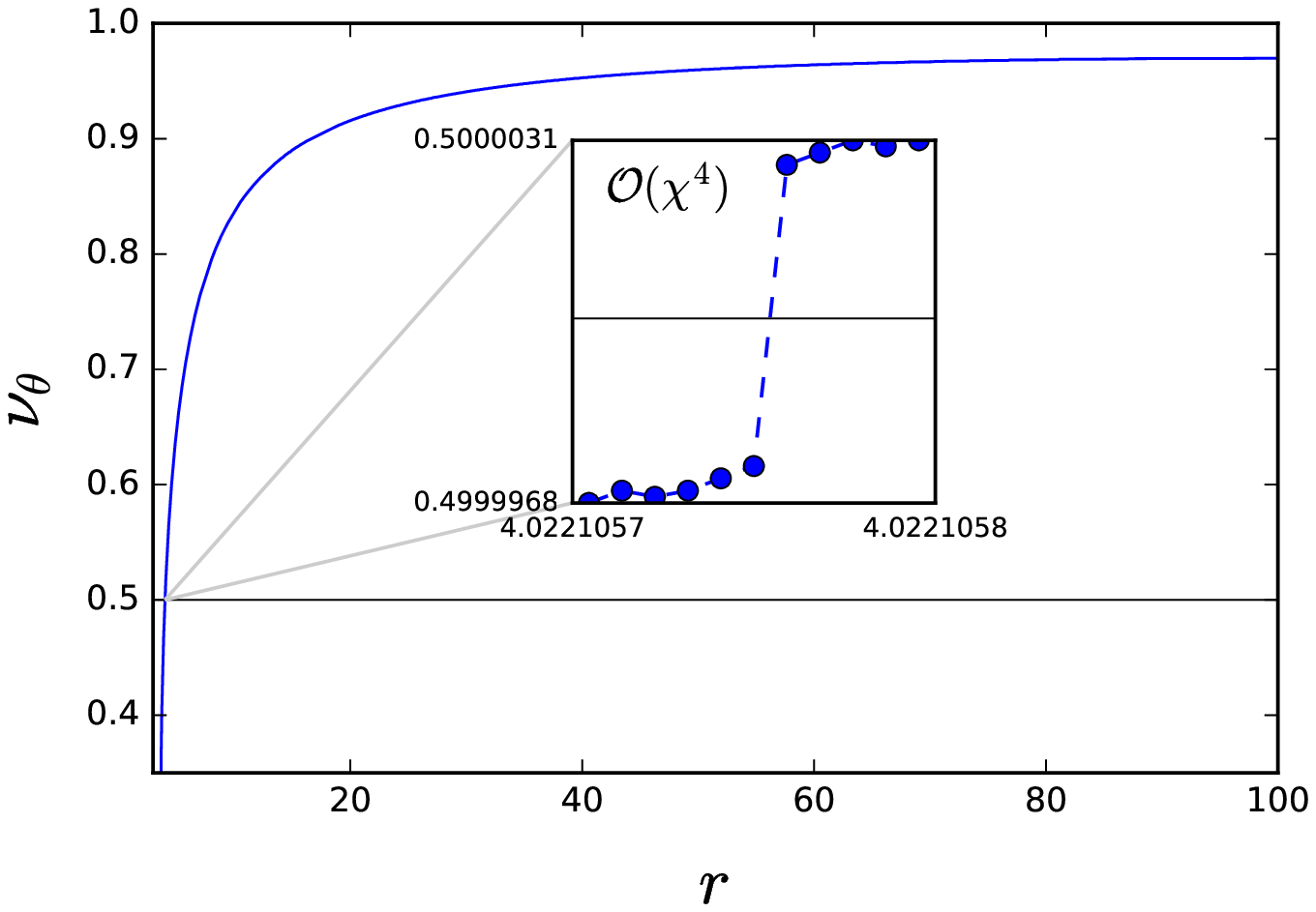} 
&
\includegraphics[width=0.53\columnwidth{}]{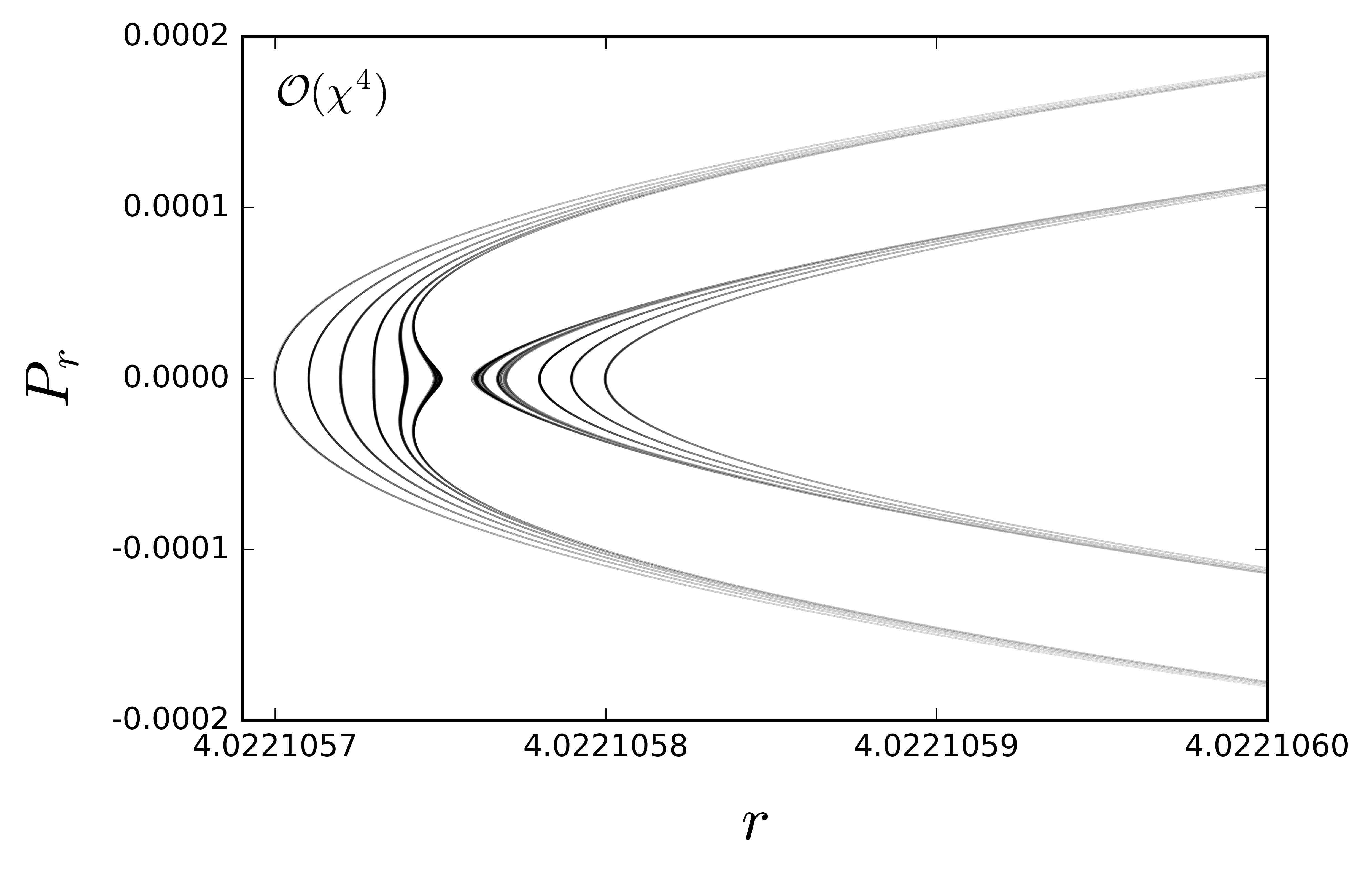} 
\end{tabular}
\end{center}
\caption{(Color Online) Rotation curves (left) and surfaces of section (right) for geodesics with $E=0.995$ and angular momentum $L_{z}=3.75365M$ around a SMBH with $\chi=0.2$, using a metric expanded to ${\cal{O}}(\chi^{4})$. Embedded in the left figure is a zoom of the rotation curve around a regime that presents chaotic features. The horizontal line is drawn at $\nu_{\theta}=0.5$. 
} 
\label{fig:RotCurves-2}%
\end{figure*}

Since we know that these Birkhoff chain of islands is caused by the slow-rotation expansion, the chain should be affected by the order of the expansion we consider. As a proof of concept, let us keep all parameters fixed, i.e., the energy, angular momentum and spin of the BH, and study the same $2/4$-resonance for a metric expanded to ${\cal{O}}(\chi^{4})$. The left panel of Fig.~\ref{fig:RotCurves-2} shows the rotation curves, which this time present an abrupt change in the rotation number, but no longer a plateau (even when the numerical precision is increased by several orders of magnitude). Zooming to this region, the right panel of Fig.~\ref{fig:RotCurves-2} shows the Poincar\'e surfaces of section near the $2/4$-resonance. The stable periodic orbits are located in the blank areas between the depicted KAM curves. To ${\cal{O}}(\chi^{4})$, therefore, the system is said to be \emph{slightly non-integrable} and numerous invariant tori survive the perturbation, albeit deformed. 

A measure for the ``amount'' of chaos in a given set of islands can be estimated by measuring the size of the plateau or the abrupt jump in the rotation number, provided that the Poincar\'e surface of section corresponds to the same system, i.e., exactly the same value for all the parameters. We find that the size of these chaotic features shrinks when the order of the approximation is increased, as expected. For example, for the cases shown in Fig.~\ref{fig:RotCurves} at ${\cal{O}}(\chi^{2})$, the size of the plateau is approximately $10^{-3}\,[r/M]$ (see the vertical dashed lines in the figure), while for Fig.~\ref{fig:RotCurves-2} at ${\cal{O}}(\chi^{4})$ the size of the abrupt change in the rotation number is smaller than $10^{-6}\,[r/M]$. 

%---------------------------------------------------------------------------------------------------------------------------------------
\subsection{Unbound Orbits}

\begin{figure*}[hpt]
\begin{tabular}{ll}
\includegraphics[width=0.5\columnwidth{}]{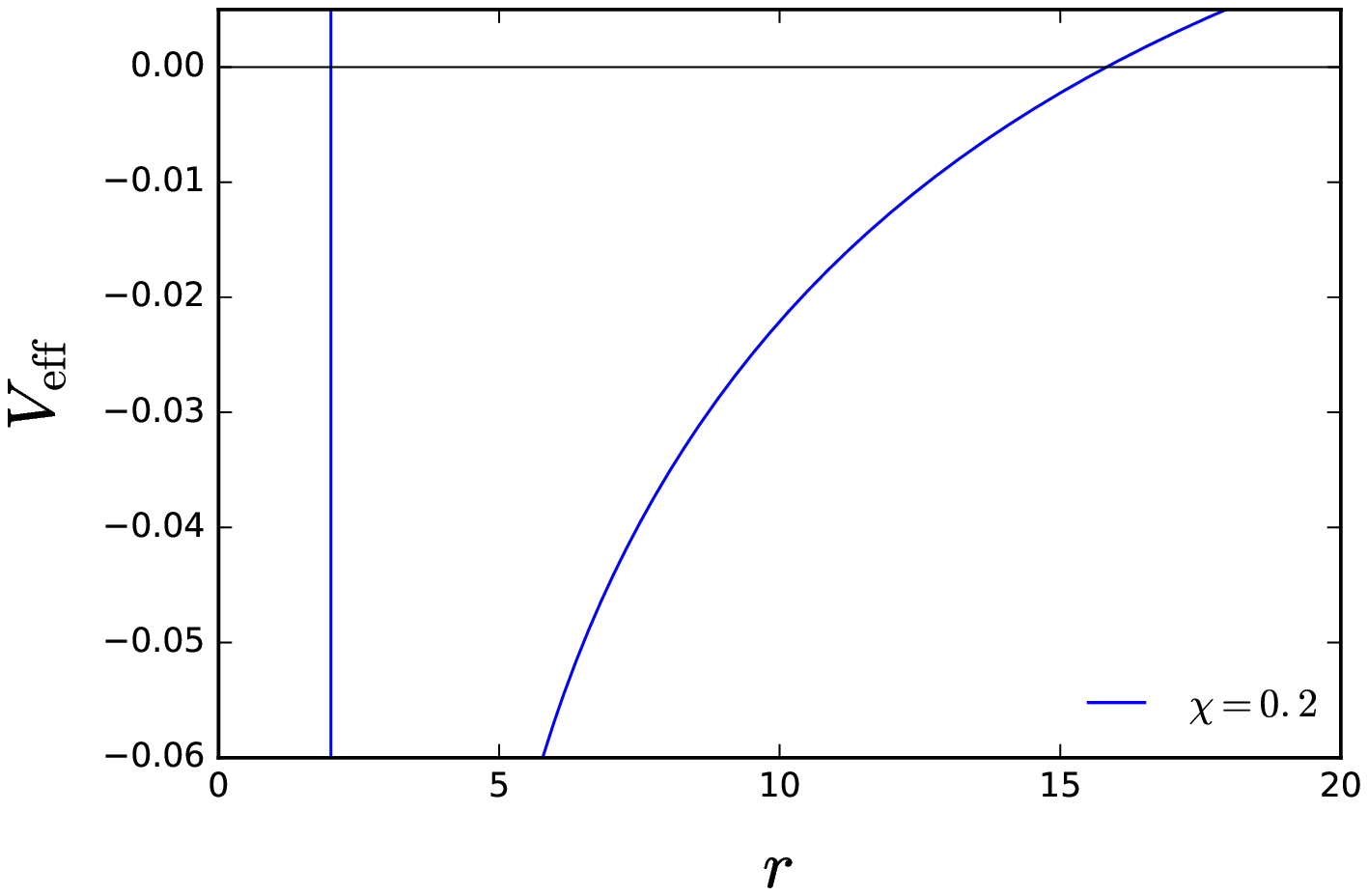}
&
\includegraphics[width=0.5\columnwidth{}]{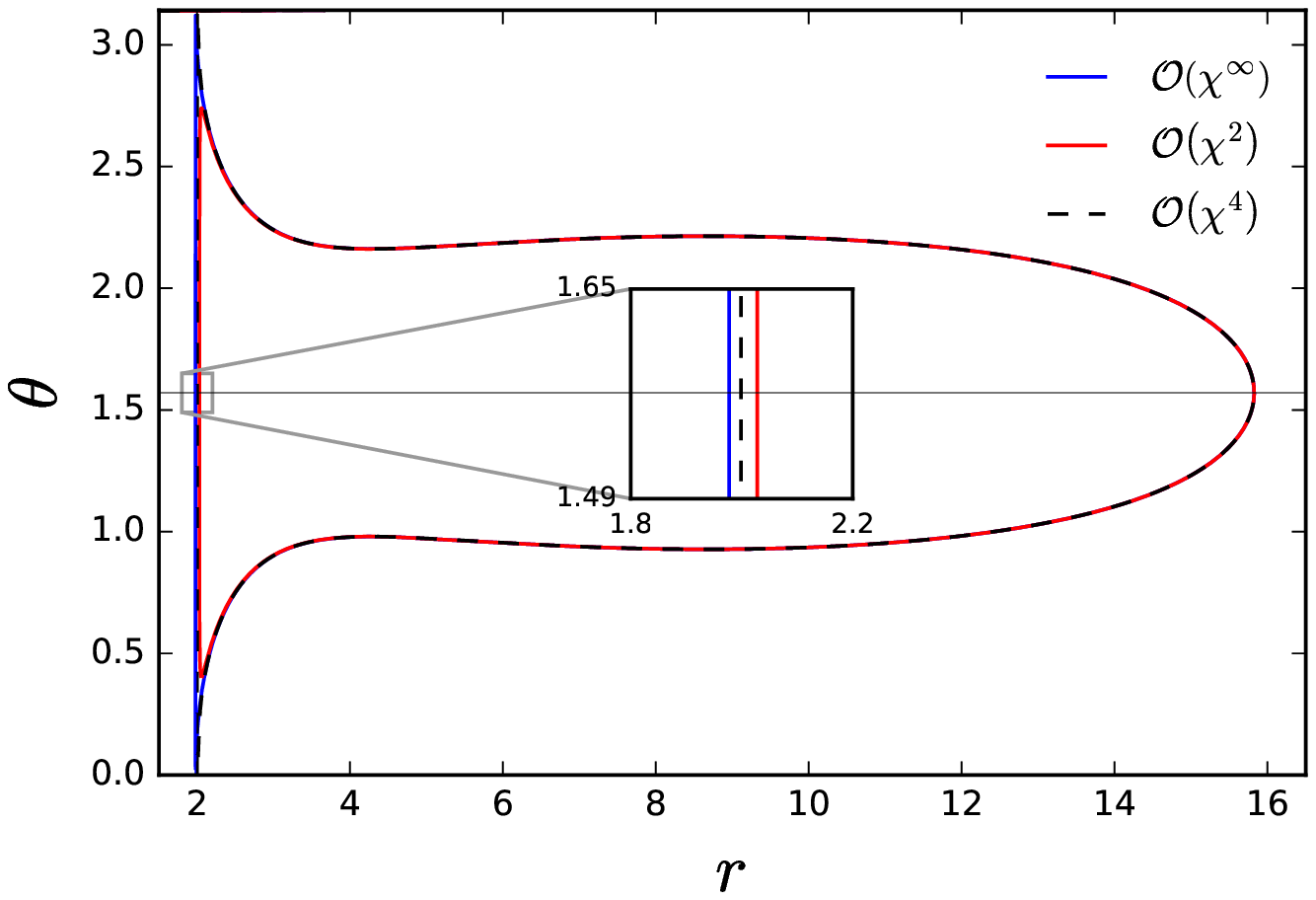} 
\end{tabular}
\caption{(Color online) Effective potential (left) and CZVs (right) using the Kerr metric in its exact form, expanded to second order and to fourth order in slow rotation, for geodesics with energy $E=0.95$ and angular momentum $L_{z}=2.85M$ around a SMBH with spin. This set of parameters allows for unbounded motion and connects the two regions that were depicted previously in Fig.~\ref{fig:Potentials}. The vertical line in the right panel shows the SMBH event horizon, which shifts slightly with the order of the approximation.
} 
\label{fig:OrderAnalysis}%
\end{figure*}

Let us now study unbounded orbits, where ultimately the small compact object plunges into the horizon of the SMBH. But instead of studying geodesics constrained to two separated regions in phase space (one for bounded orbits and one for unbounded ones) as considered in Fig.~\ref{fig:Potentials}, let us study geodesics that can communicate between these regions. One can choose values of $E$, $L_{z}$ and $\chi$ such that these two regions are connected, as shown in Fig.~\ref{fig:OrderAnalysis} through the effective potential and its respective CZVs for metrics truncated at different orders. For such geodesics, certain regions of the surfaces of section present heteroclinic chaos (recall that this is caused by geodesics that visit the same equilibrium point, see Sec.~\ref{Framework}), which lead to stronger features of chaos~\cite{LukesGerakopoulos:2010rc}. 

\begin{figure*}[hpt]
\begin{tabular}{ll}
\includegraphics[width=0.5\columnwidth{}]{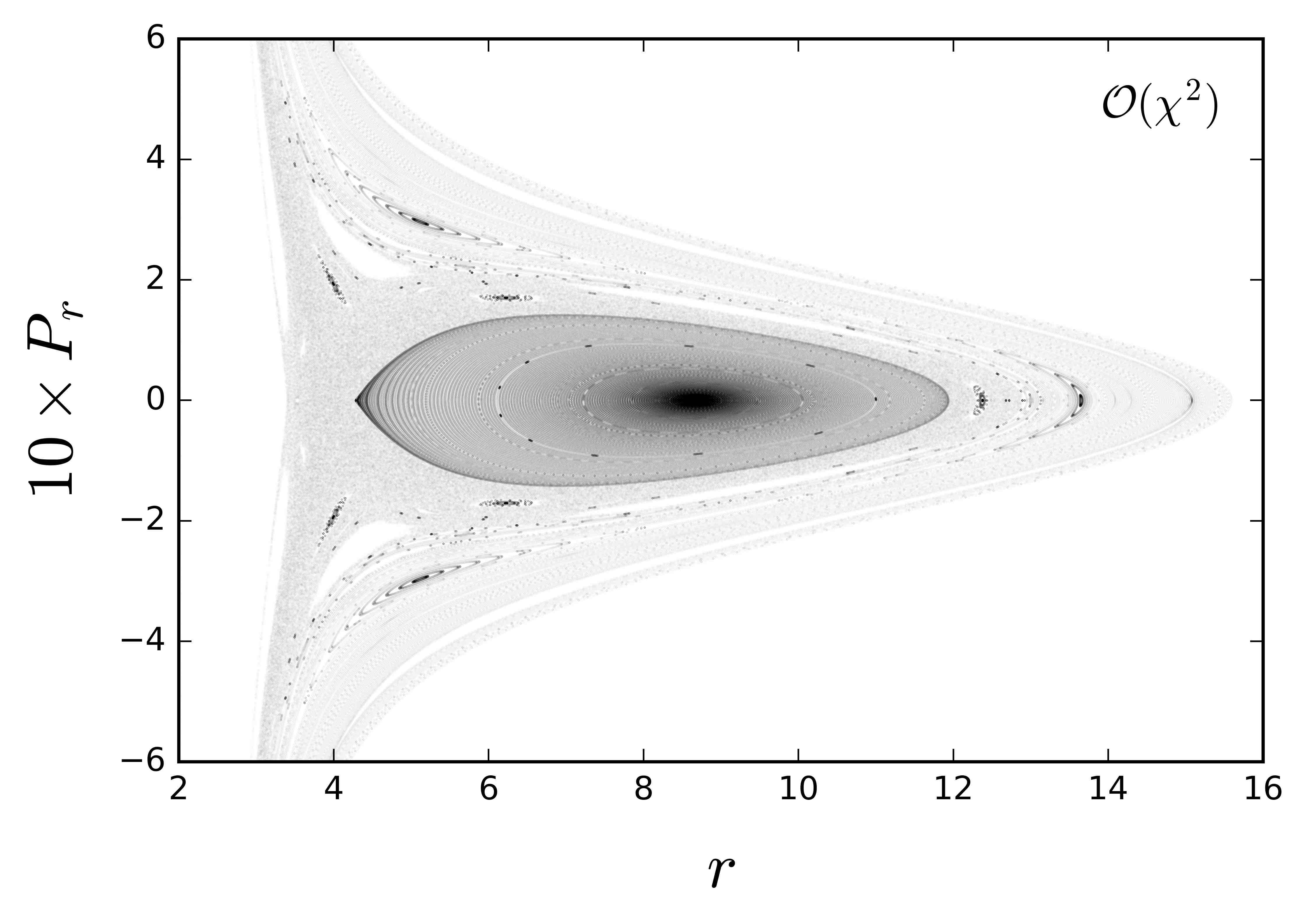}
&
\includegraphics[width=0.5\columnwidth{}]{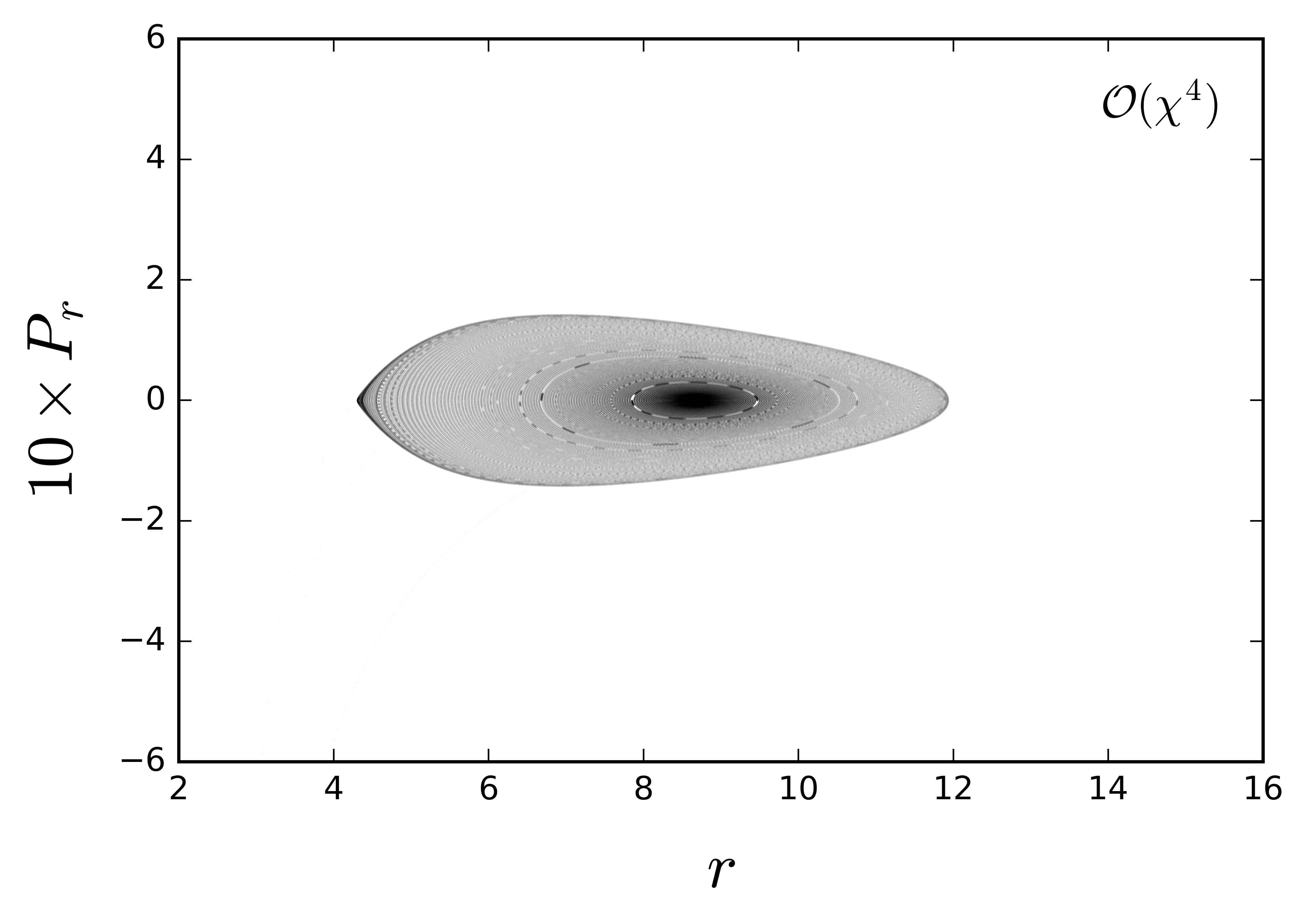} 
\end{tabular}
\caption{Poincar\'e surfaces surface of section for  $\chi=0.2$, $E=0.95$ and angular momentum $L_{z}=2.85M$ for (left) the metric expanded to ${\cal{O}}(\chi^{2})$ and (right) to ${\cal{O}}(\chi^{4})$. See text for more details.} 
\label{fig:SeaChaosIslands}%
\end{figure*}

Given that we expect stronger signatures of chaos, let us jump directly to a study of the Poincar\'e surfaces of section for unbounded orbits. The left panel of Fig.~\ref{fig:SeaChaosIslands} shows the surfaces of section when using the metric to ${\cal{O}}(\chi^{2})$. Observe the main island in the center of the figure, surrounded by a chaotic sea of layers with many high-multiplicity islands of stability; most of the chaotic orbits correspond to plunging orbits. On the other hand, the surfaces of section when using the metric to ${\cal{O}}(\chi^{4})$ show none of this manifestly chaotic behavior. In fact, the chaotic sea disappears completely even when the resolution is increased, as shown in the right panel of Fig.~\ref{fig:SeaChaosIslands}. Nevertheless, near last KAM curve there should indeed exist chaotic orbits plunging to the central object, but the size of the region where they occurred has  decreased significantly. This reinforces the findings of the previous section, i.e., that the chaotic behavior we found scales with the order of the slow-rotation approximation. 
%

%%%%%%%%%%%%%%%%%%%%%%%%%%%%%%%%%%%%%
\section{Geodesics in dynamical Chern-Simons}
\label{GeodesicsdCS}
%%%%%%%%%%%%%%%%%%%%%%%%%%%%%%%%%%%%%%

The modification introduced by dCS to the Kerr solution is small, in a perturbative sense, and the equations of motion are still separable after orbiting averaging, except at resonant orbits~\cite{Yagi:2012ya}. Nevertheless, there could be secular changes in the angular frequencies $\omega^{\mu}$ and a second rank Killing tensor, associated with a Carter-like constant, could not be found in Ref.~\cite{Yagi:2012ya}. The lack of an exact solution in dCS valid for all spin magnitudes forces us to question the regime of validity of the metric, whether the approximate nature of the spacetime has a significant impact on possible observables, and if it is related to the appearance of chaotic behavior. These are the topics we will study in this section.

We employ two different dCS metrics: an expanded metric and a resummed metric. The expanded metric is 
\be
g_{\mu \nu}^{\expp} = g_{\mu \nu}^{\SRKerr,n} + g_{\mu \nu}^{\dCS,n}\,,
\label{eq:exp-metric}
\ee
where $g_{\mu \nu}^{\SRKerr,n}$ is the Kerr metric expanded to ${\cal{O}}(\chi^{n})$ in $\chi \ll 1$ and $g_{\mu \nu}^{\dCS,n}$ is the dCS correction presented in Eq.~(\ref{solution}) and in~\ref{app:ExplicitMetric}, which is an expansion in $\chi \ll 1$ also to ${\cal{O}}(\chi^{n})$ and in $\zeta \ll 1$ to ${\cal{O}}(\zeta^{1})$. The resummed metric is 
\be
g_{\mu \nu}^{\resum} = g_{\mu \nu}^{\Kerr} + g_{\mu \nu}^{\dCS,n}\,,
\label{eq:resum-metric}
\ee
where $g_{\mu \nu}^{\dCS,n}$ is the same as in the expanded metric, but $g_{\mu \nu}^{\Kerr}$ is the exact Kerr metric. In all cases, we always employ Boyer-Lindquist coordinates. In fact, the transformation of the dCS metric from Hartle-Thorne coordinates to Boyer-Lindquist coordinates is what allows us to resum the $\zeta$-independent sector, as the Hartle-Thorne metric is intrinsically defined in a slow-rotation expansion.

The first topic we investigate is the effect of the coupling parameter $\zeta$ at a fixed order in the slow-rotation expansion on chaotic features of bounded geodesics. For this study, we focus on the rotation number and employ the expanded dCS metric of Eq.~(\ref{eq:exp-metric}) truncated to ${\cal{O}}(\chi^{2})$. As expected, we find plateaus in the rotation number, just as we did in the slow-rotation expansion of Kerr discussed in Sec.~\ref{sec:bounded-orbits}. Table~\ref{table:PlateauSizedCS} shows the size of the plateaus for geodesics with the same parameters as those of Fig.~\ref{fig:RotCurves}. Observe that the plateau size remains roughly constant and the effect of $\zeta$ is very small. We find that generically, for bounded orbits, increasing $\zeta$ decreases the size of the plateau\footnote{The decrease of size of the plateau is because for the parameters chosen when $\zeta$ increases, the well of the effective potential that allows for bounded orbits is shifted away from the source.}. These results suggest that the slow-rotation expansion may be responsible for the appearance of these chaotic signatures in dCS gravity. 

\begin{table}[htb]
\caption{Size of the plateau in the rotation number for geodesics with $E=0.995$ and angular momentum $L_{z}=3.75365M$ and a SMBH spin of  $\chi=0.2$, using the expanded dCS metric to ${\cal{O}}(\chi^{2})$ and three different values of the coupling parameter $\zeta$. The effect of $\zeta$ on the plateau size is small.}
\begin{center}
\begin{tabular}{|c|c|}
\hline 
$\zeta$ & Plateau's size $[r/M]$ \tabularnewline
\hline 
\hline 
$0.00$ & $0.00119$\tabularnewline
\hline 
$0.05$ & $0.00116$\tabularnewline
\hline 
$0.10$ & $0.00114$\tabularnewline
\hline 
$0.20$ & $0.00110$\tabularnewline
\hline 
\end{tabular}
\par
\end{center}
\label{table:PlateauSizedCS}%
\end{table}

\begin{figure*}[hpt]
\begin{center}
\begin{tabular}{ll}
\includegraphics[width=0.5\columnwidth{}]{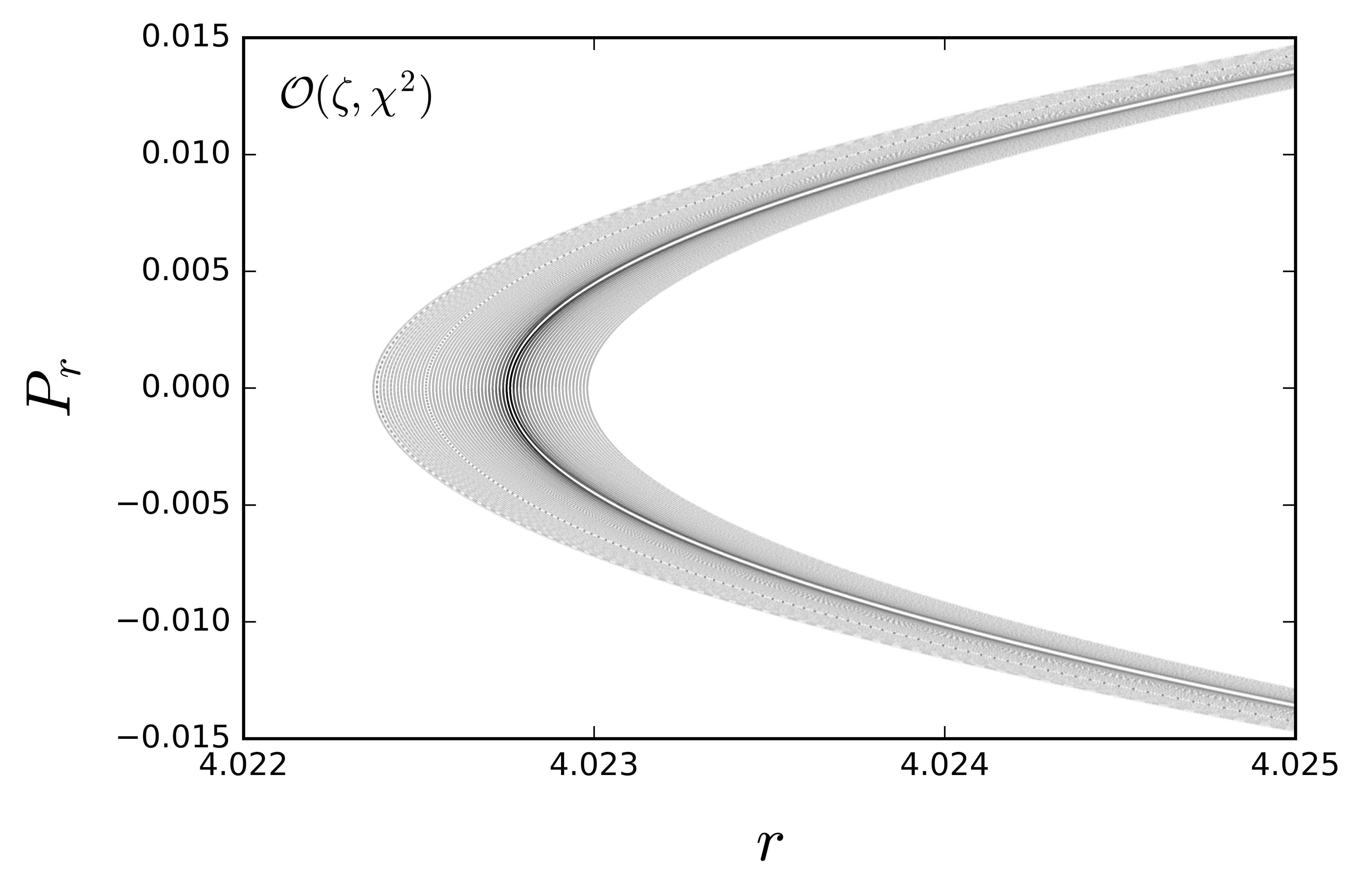}
&
\includegraphics[width=0.5\columnwidth{}]{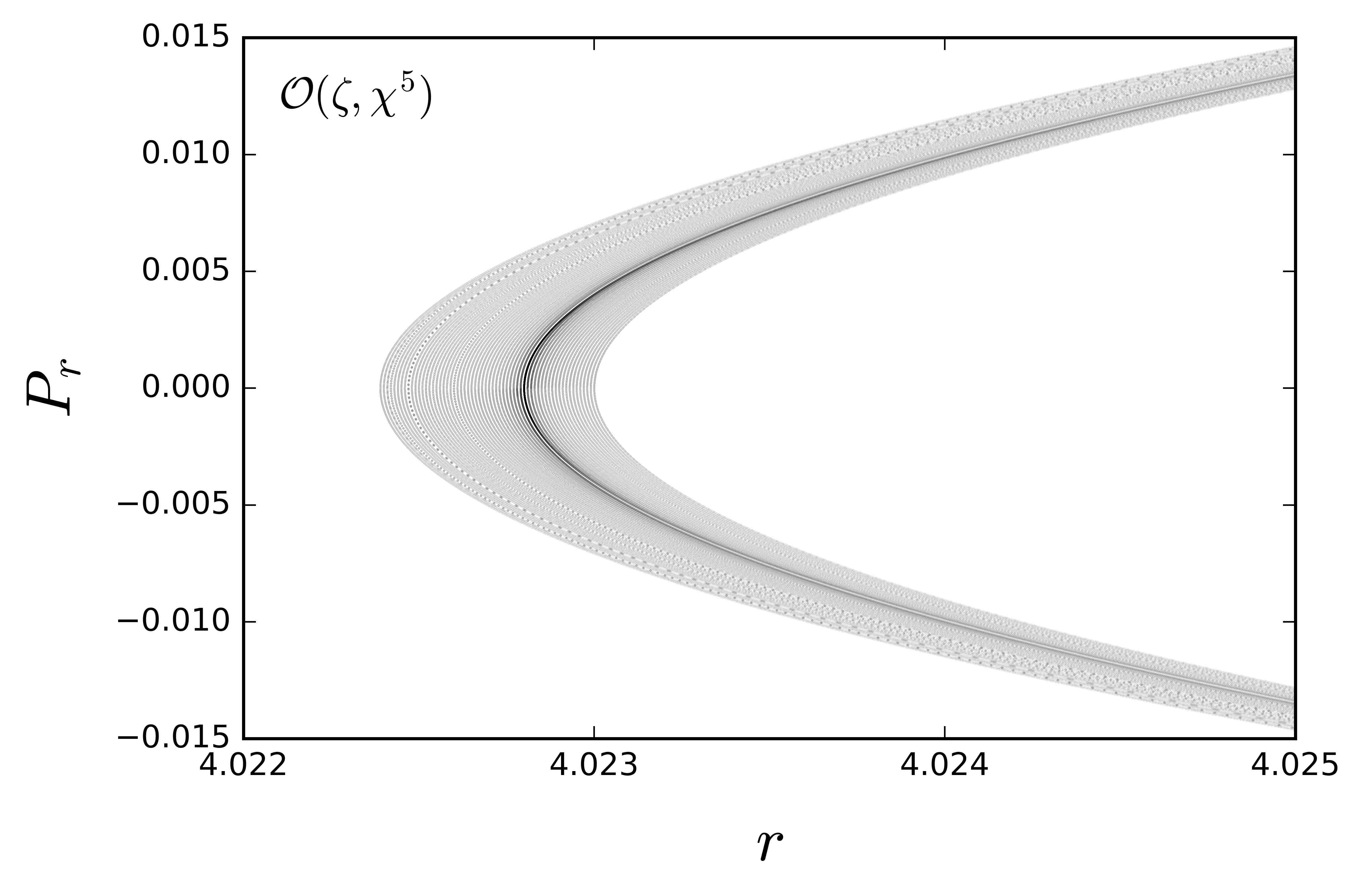} 
\end{tabular}
\end{center}
\caption{The surface of section for $\chi=0.2$, $\zeta=0.2$, $E=0.995$ and angular momentum $L_{z}=3.75365M$ for (left) the metric to ${\cal{O}}(\chi^{2})$ and (right) to ${\cal{O}}(\chi^{5})$. We see that the region of the KAM curves decrease considerably when the order of the approximation increases, see the scales in the radius $r$.} 
\label{fig:SectionsCS}%
\end{figure*}

Let us study this hypothesis by calculating the surfaces of section with the resumed metric. Figure~\ref{fig:SectionsCS} shows these surfaces for the same geodesic and SMBH parameters used in Fig.~\ref{fig:RotCurves}, fixing $\zeta=0.2$ and using both a dCS correction to ${\cal{O}}(\chi^2)$ (left panel) and to ${\cal{O}}(\chi^5)$ (right panel). Observe that the tori structure in phase space is deformed, as we found when using the slow-rotation expansion of the Kerr metric to high order in $\chi$. Moreover, observe that the deformation decreases with the order in $\chi$ kept in the dCS deformation. All of this suggests that the resumed dCS metric leads to a slightly non-integrable system where most of the invariant tori survive the dCS perturbation, albeit deformed. 

\begin{figure*}[hpt]
\begin{center}
\begin{tabular}{ll}
\includegraphics[width=0.5\columnwidth{}]{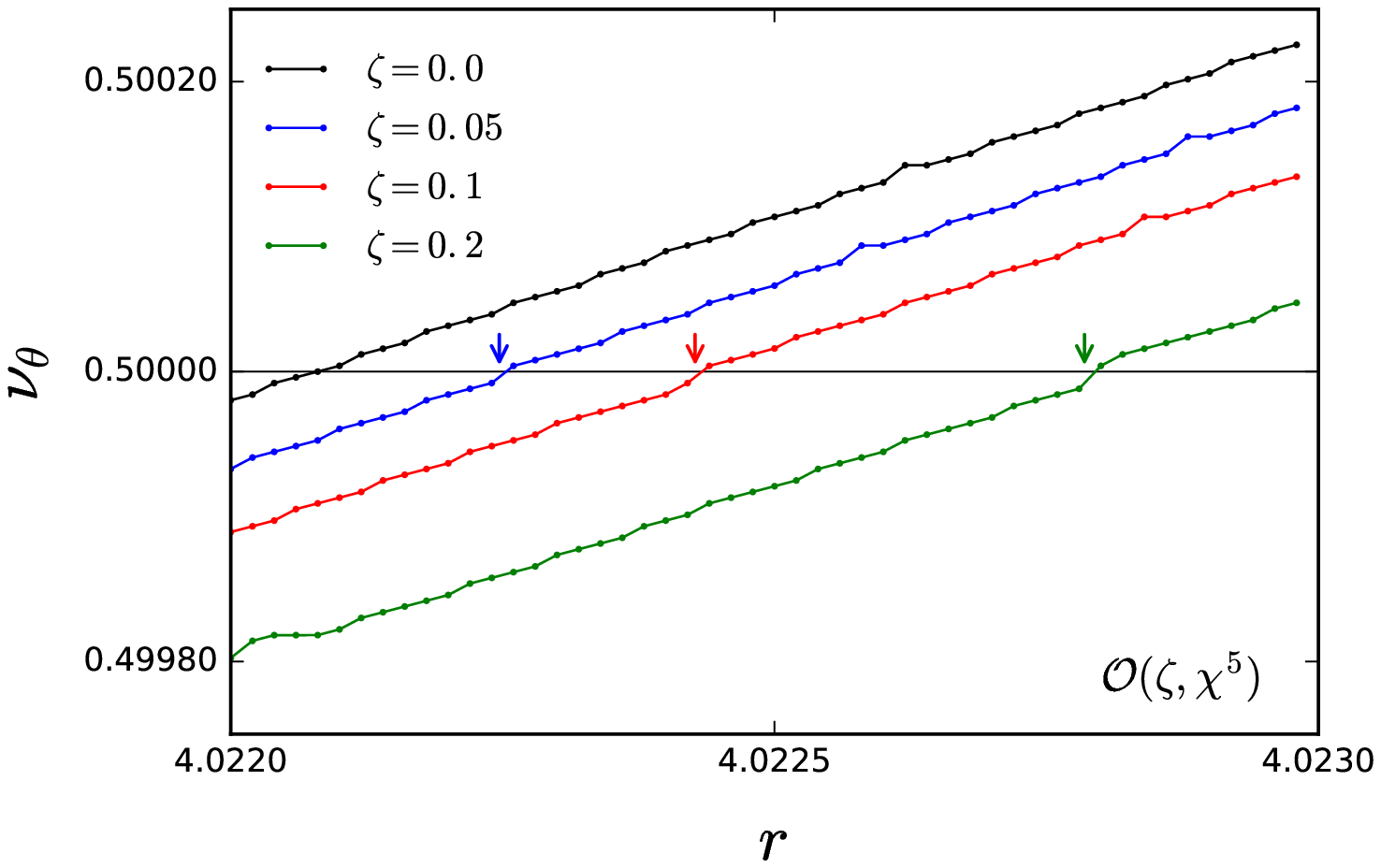}
&
\includegraphics[width=0.5\columnwidth{}]{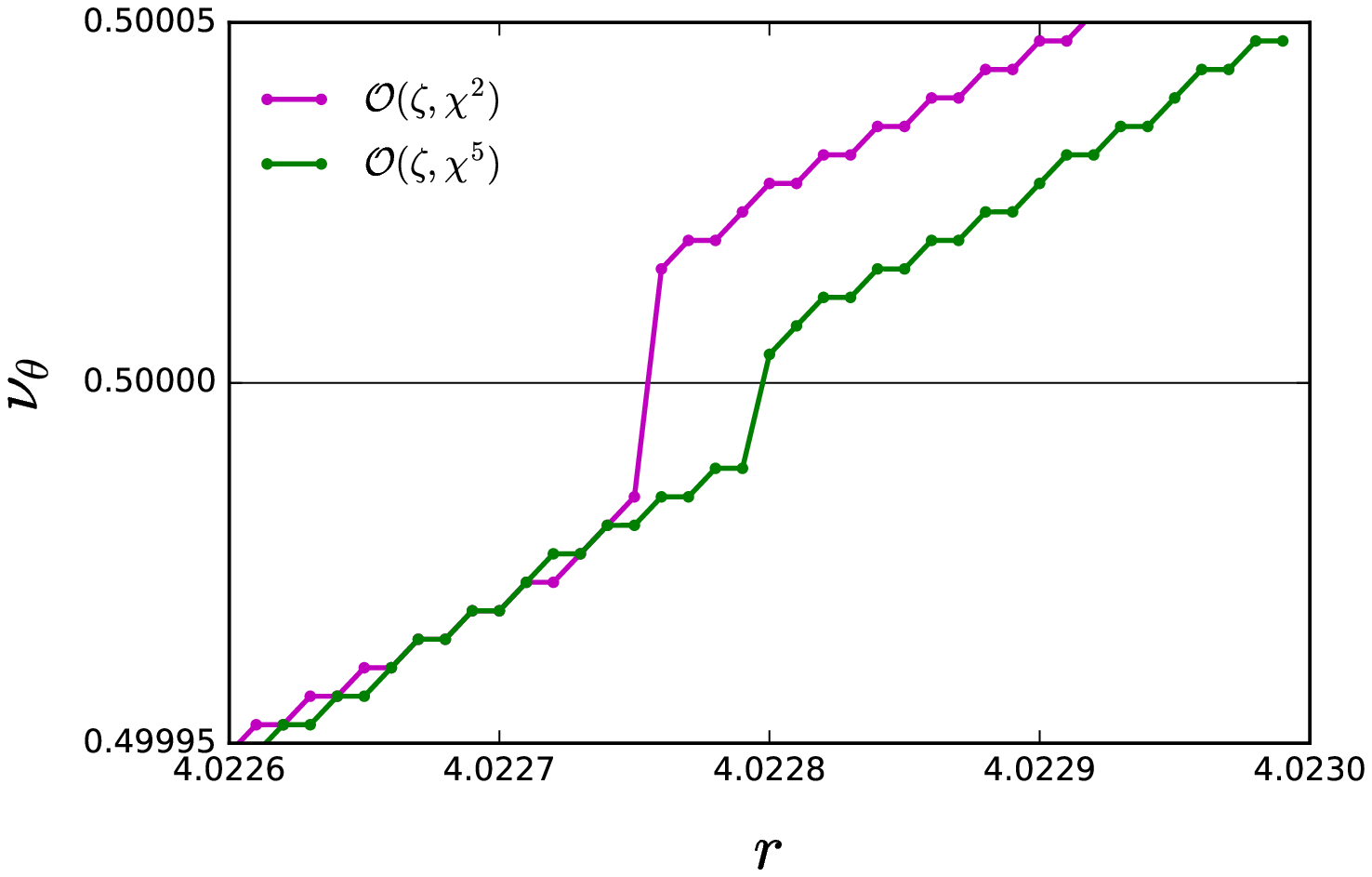} 
\end{tabular}
\end{center}
\caption{(Color Online) Rotation curves for geodesics with $E=0.995$ and angular momentum $L_{z}=3.75365M$ around a SMBH with $\chi=0.2$, using the resumed dCS metric truncated at ${\cal{O}}(\chi^{5})$ (left) for different values of $\zeta=0.0,0.05,0.1,0.2$ and truncated at ${\cal{O}}(\chi^{2},\zeta)$ and ${\cal{O}}(\chi^{5},\zeta)$ with $\zeta=0.2$ (right). Observe in the left panel that the abrupt jump in the rotation number is insensitive to the value of $\zeta$. Observe also in the right panel that the abrupt jump decreases with the order in $\chi$ kept in the dCS correction.} 
\label{fig:RotCurvedCS}%
\end{figure*}

Given that chaotic features could be hiding in the surfaces of section at sufficiently high resolution, let us now study the rotation curves of the geodesics in the resumed dCS spacetime. The left panel of Fig.~\ref{fig:RotCurvedCS} shows various rotation curves calculated using different values of $\zeta$, for the same geodesics of Fig.~\ref{fig:SectionsCS}. Observe that when $\zeta = 0$, no features in the rotation curve are observed, and when $\zeta \neq 0$, although an abrupt jump appears, no plateau emerges. Moreover, as in the case of the expanded metric, the radial size of the jump is approximately constant with $\zeta$. This suggests that the abrupt jump in the rotation number is related to the truncation of the dCS correction in $\chi$ and is not a feature of the dCS spacetime. 

Let us investigate this conclusion by fixing the value of $\chi$ and changing the order of truncation of the dCS correction. The right panel of Fig.~\ref{fig:RotCurvedCS} shows the abrupt jump in the rotation curve computed with the resumed dCS metric and the dCS correction truncated at different orders in $\chi$. Observe that as higher order in $\chi$ terms are kept in the dCS correction, the size of the abrupt jump decreases. These results reinforce the hypothesis that the abrupt jump is caused by the truncation of the dCS correction in $\chi$. 

\begin{figure*}[hpt]
\begin{tabular}{ll}
\includegraphics[width=0.5\columnwidth{}]{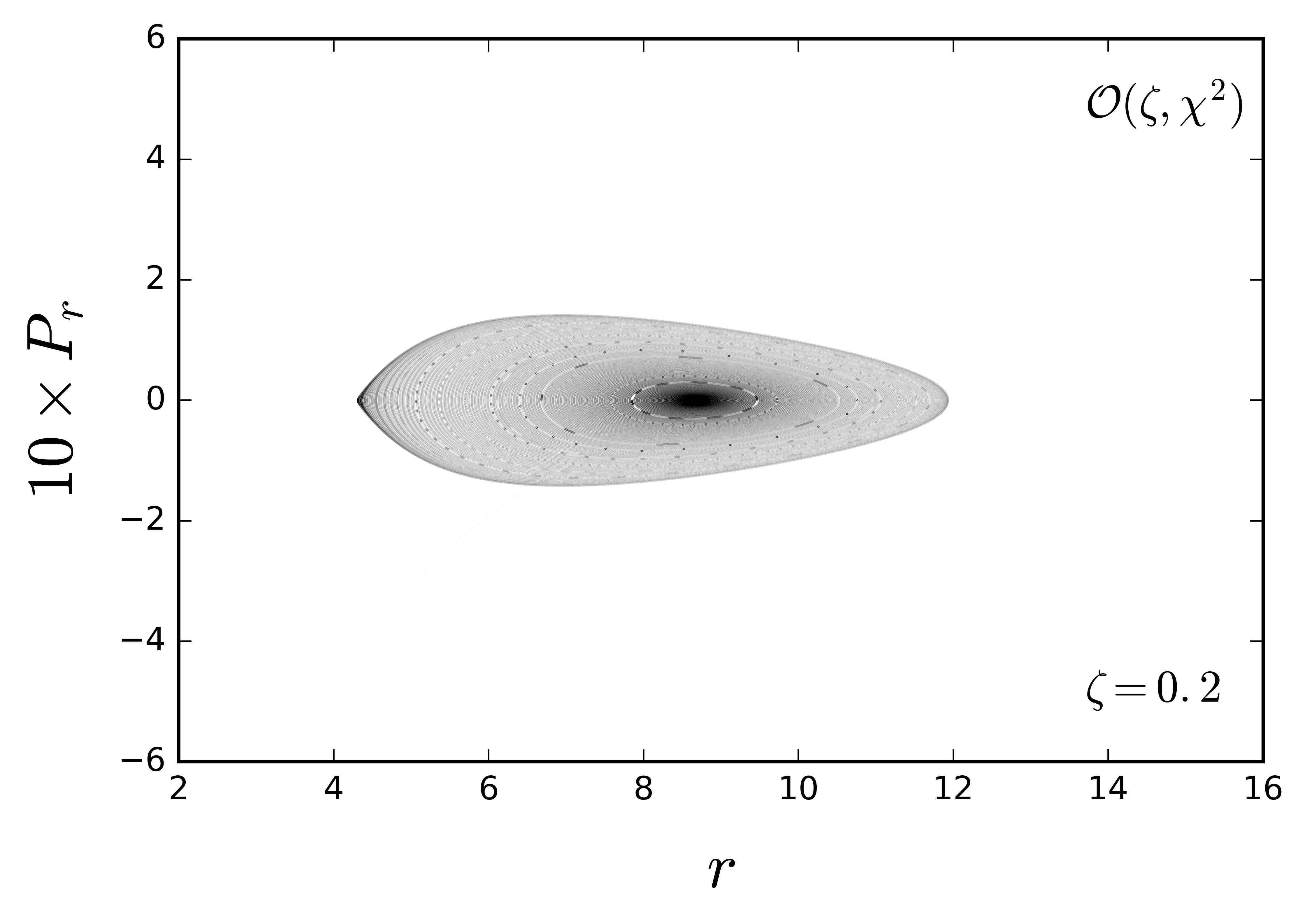}
&
\includegraphics[width=0.5\columnwidth{}]{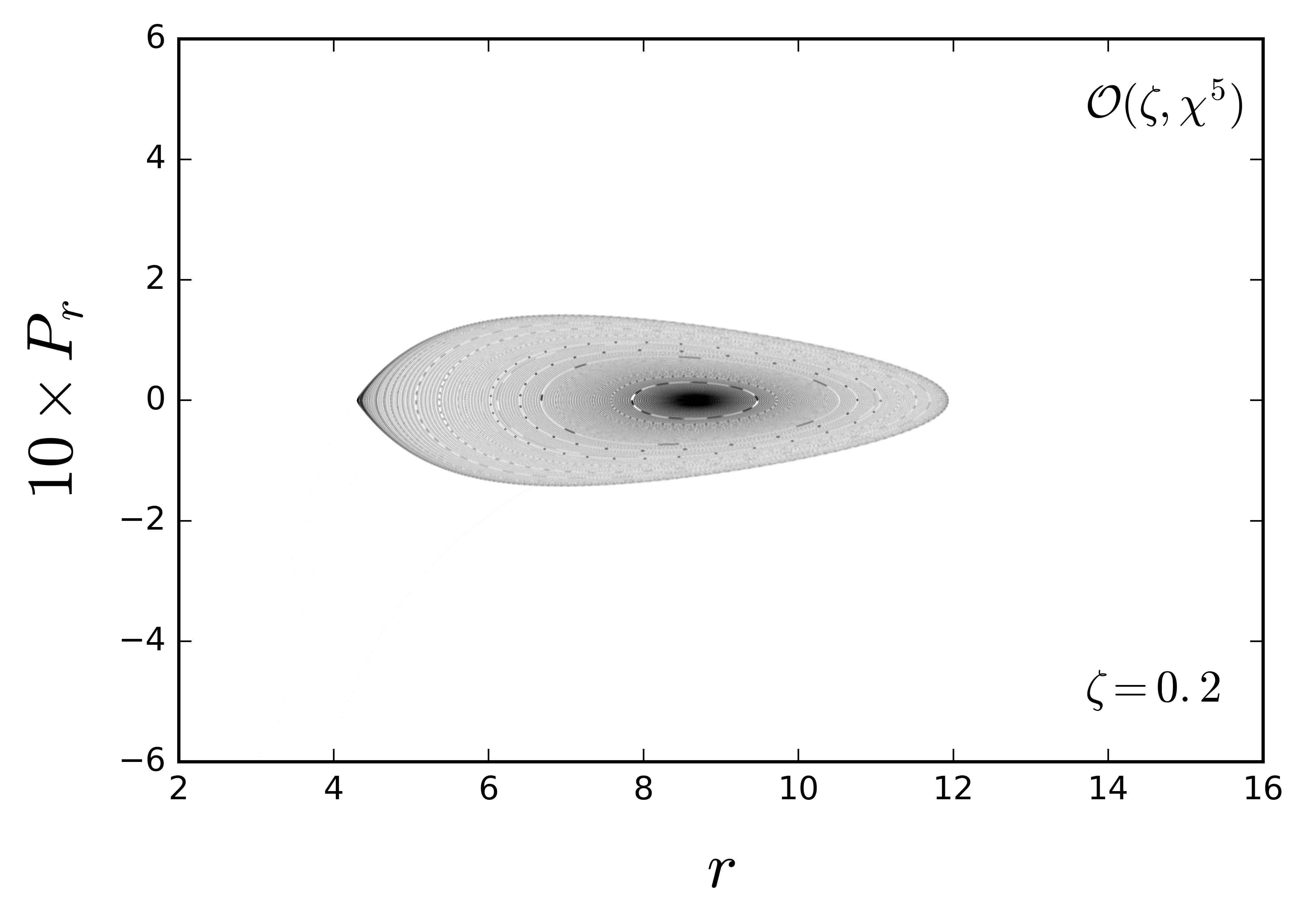} 
\end{tabular}
\caption{Poincar\'e surfaces of section for  $\zeta=0.2$, $\chi=0.2$, $E=0.95$ and angular momentum $L_{z}=2.85M$ for (left) the metric expanded to ${\cal{O}}(\zeta, \chi^{2})$ and (right) to ${\cal{O}}(\zeta,\chi^{5})$.} 
\label{fig:NonConDCS}%
\end{figure*}

We also investigated unbounded orbits, but unlike in the slowly-rotating Kerr case, the resummed dCS metric did not produce any clear signatures of chaos, even for relatively high values of $\zeta$. In Fig.~\ref{fig:NonConDCS}, we show Poincar\'e surfaces of section using the resummed metric with the dCS piece truncated at (left panel) ${\cal{O}}(\chi^{2})$ and at (right panel) ${\cal{O}}(\chi^{5})$. This figures are to be compared with Fig.~\ref{fig:SeaChaosIslands}, which we recall presents surfaces of section using a slowly-rotating Kerr metric. Observe that Fig.~\ref{fig:NonConDCS} does not contain the sea of chaos that is clearly visible in Fig.~\ref{fig:SeaChaosIslands}.

All of these results suggest that geodesics in dCS gravity are not chaotic. Proving such a statement, of course, is not possible through a purely numerical analysis, since one could always imagine that chaotic features arise at smaller scales than those proved by our numerics, or perhaps for another set of geodesic parameters we did not consider. To address the first point, we carried out a detailed numerical analysis at various levels of resolution and with different integration routines; in all cases we find the same results as those presented above. To address the second point, we carried out an extensive numerical analysis for a large set of geodesic parameters; in all cases we again find the same results as the representative examples discussed above. Given all of this, we conclude that geodesics in dCS gravity are not chaotic if one had an \emph{exact} dCS black hole solution and are at worst slightly non-integrable with thin chaotic layers, that scale in size with the order of the approximation when using the resummed dCS metric, surrounded by deformed invariant tori.

%%%%%%%%%%%%%%%%%%%%%%%%%%%%%%%%%%%%%%%%%%%%%
%%%%%%%%%%%%%%%%%%%%%%%%%%%%%%%%%%%%%%%%%%%%%
\section{Discussion and Concluding Remarks}
\label{disc}

We have investigated extreme mass-ratio inspirals in dCS gravity through a test-particle approximation to determine whether chaotic motion emerges in this theory. We began through an analysis of geodesics in the Kerr spacetime using both an exact Kerr metric and a slow-rotation expansion. The use of the latter is important because the dCS metric of a spinning BH is only known in the slow-rotation approximation. We found that chaotic features arise for geodesics of a slow-rotation expansion of the Kerr metric, but these chaotic features effectively disappear as higher order terms in the expansion are kept in the metric. We then studied geodesics in a dCS spacetime, using both an expanded metric and a resummed metric, where all dCS independent terms are collected to resum the Kerr metric. We found that geodesics in the resummed dCS spacetime are slightly non-integrable with thin chaotic layers, that scale in size with the order of the approximation when using the resummed dCS metric. We expect that the family of tori recovers its continuity for the exact solution and we conjecture that geodesics of the exact spinning dCS BH metric are not chaotic.

Our numerical findings also suggest that geodesics in an \emph{exact} dCS background may have a fourth constant of the motion. Reference~\cite{Yagi:2012ya} showed that at second order in rotation there does not exist a 2nd-rank Killing tensor, and thus, that a Carter-like constant of motion does not exist. Moreover, Ref.~\cite{Yagi:2012ya} also showed that there does not exist a coordinate system in which the second-order-in-rotation metric satisfies the Levi-Civita test~\cite{levi1904sulla,Yasui:2011pr}, which implies that the Hamilton-Jacobi equations are not additively separable through the existence of a 2nd-rank Killing tensor. These results, however, do not imply that a higher-rank Killing tensor does not exist, and thus, they do not rule out the existence of a fourth-constant of motion associated with such a higher-rank Killing tensor, or the existence or not of chaos.  Therefore, the results of Ref.~\cite{Yagi:2012ya} are \emph{not} in conflict with our findings, and in particular, with our conjecture, i.e., that the not-yet-found dCS metric valid to all orders in spin does not lead to chaotic geodesics, which then implies the existence of a 4th constant of the motion associated with a higher-rank Killing tensor. These findings have implications on ongoing efforts to find an exact solution in dCS gravity for spinning BHs.

The analysis we have carried out, however, is only an approximation to the motion of extreme mass-ratio inspirals in dCS gravity. Needless to say, we have not included dissipative effects in the orbits, which is why we were able to study bounded motion. The inclusion of such dissipative effects with the resummed dCS metric could be the topic of future studies. Furthermore, we have here neglected non-geodesic forces induced by the scalar (magnetic dipole) charge of the small object. Such a force may lead to interesting corrections on the motion found here that could also be studied further. 

Ultimately, once such studies have been carried out, one can investigate the signatures of dCS gravity on extreme mass-ratio inspirals. The only studies on this topic are those of~\cite{Sopuerta:2009iy,Canizares:2012is}, which did not considered a resummed metric and did not include the magnetic-dipole force due to the small object. Redoing such an analysis would allow us to estimate the accuracy to which dCS gravity could be constrained by future gravitational wave observations with LISA. 

%%%%%%%%%%%%%%%%%%%%%%%%%%%%%%%%%%%%%%%%%%%%%
%%%%%%%%%%%%%%%%%%%%%%%%%%%%%%%%%%%%%%%%%%%%%

\section*{Acknowledgements}

We thank Neil Cornish, Dimitry Ayzenberg, Kent Yagi, Georgios Loukes-Gerakopoulos, Paolo Pani, Andrea Masselli, Charles Kankelborg, Thomas Sotiriou, Olivier Sarbach and Houri Tsuyoshi for very useful suggestions and comments. A.C.-A. acknowledges funding from the Fundaci\'on Universitaria Konrad Lorenz (Project 5INV1). N.Y. and A.C.-A. acknowledge financial support through NSF CAREER grant PHY-1250636 and NASA grants NNX16AB98G and 80NSSC17M0041. L.A.P. acknowledges funding from the Comit\'e para el Desarrollo de la Investigaci\'on -CODI- of Universidad de Antioquia, Colombia under the Estrateg\'ia de Sostenibilidad. Computational efforts were performed on the Hyalite High-Performance Computing System, operated and supported by Montana State University's Information Technology Center.

%%%%%%%%%%%%%%%%%%%%%%%%%%%%%%%%
\appendix

%%%%%%%%%%%%%%%%%%%%%%%%%%%%%%%%

\section{BH Solutions in dCS gravity to fifth order in Boyer Lindquist Coordinates}
\label{app:ExplicitMetric}

The dCS metric in Boyer-Lindquist coordinates $(t,r,\theta,\phi)$ is obtained from the solution given in Ref.~\cite{Maselli:2017kic} in Hartle-Thorne coordinates $(t,r_{\HT},\theta_{\HT},\phi)$, after making a coordinate transformation implicitly given by,

\small
\begin{eqnarray}
\label{eqn:Transformation}
r_{\HT} &=& \chi ^4  \frac{M^4}{40 r^7}  \Bigg[ -8 M^4+72 M^3 r+36 M^2 r^2-15 M r^3-5 r^4 \nonumber \\  & & +10 \cos ^2(\theta ) (3 M+r) \left(12 M^3-18 M^2 r+r^3\right) \nonumber \\ 
& & - 5 \cos ^4(\theta ) \left(192 M^4-60 M^3 r-28 M^2 r^2+3 M r^3+r^4\right) \Bigg]\nonumber \\ & & + \frac{ M^6 \cos (2 \theta ) (3 M+r)}{602112 r^{12}} \zeta  \chi ^4   \Bigg[ 2 \left( 5757696 M^6+1656648 M^5 r \right. \nonumber \\
& & \left. -275748 M^4 r^2-709360 M^3 r^3-173364 M^2 r^4\right. \nonumber \\
& & \left. -14154 M r^5-4627 r^6 \right) +3 (2 M-r)\cos (2 \theta ) \left( 2878848 M^5 \right. \nonumber \\
& & \left. + 553140 M^4 r+226560 M^3 r^2-144380 M^2 r^3 \right. \nonumber \\
& & \left. -1092 M r^4-273 r^5 \right)
\Bigg]\,
\\ 
\theta_{\HT} &=& \frac{M^4 \sin (\theta ) \cos (\theta )}{8 r^7} \chi ^4 \Bigg[ -3 r (2 M+r)^2 \nonumber \\ 
& & +2 \cos ^2(\theta ) \left(-12 M^3+2 M^2 r+4 M r^2+r^3\right) \Bigg]\,.
\end{eqnarray}

The extra terms of the metric~(\ref{solution}) are, to $\mathcal{O}(\zeta,\chi^5)$, 

%\right. \nonumber \\ & & \left.
\small
\begin{eqnarray}
g_{tt}^\dCS &=& \zeta \chi^4  \Bigg[ \frac{1}{384} \frac{M^5}{r^5}  \left(1+\frac{2624}{35} \frac{M}{r}+\frac{492831}{3920} \frac{M^2}{r^2}+\frac{1771487}{1680} \frac{M^3}{r^3} \right.  \nonumber \\ 
& & \left. +\frac{330775}{168} \frac{M^4}{r^4} +\frac{4430511}{980} \frac{M^5}{r^5} - \frac{6957813}{980} \frac{M^6}{r^6} + \frac{6488861}{980} \frac{M^7}{r^7}\right.  \nonumber \\ 
& & \left. +\frac{667071}{70} \frac{M^8}{r^8} +\frac{15984}{5} \frac{M^9}{r^9} \right) - \frac{1819}{56448} \frac{M^3}{r^3} \left(3 \cos ^2(\theta )-1\right) \left(1+\frac{M}{r}  \right.  \nonumber \\ 
& &  \left. +\frac{51806}{12733} \frac{M^2}{r^2} +\frac{135383}{63665} \frac{M^3}{r^3} + \frac{309664}{38199} \frac{M^4}{r^4} - \frac{36264049 }{381990} \frac{M^5}{r^5}\right.  \nonumber \\ 
& &  \left.  - \frac{7873793}{38199} \frac{M^6}{r^6} - \frac{32533551}{63665} \frac{M^7}{r^7} + \frac{73025558 }{63665} \frac{M^8}{r^8} - \frac{8708988}{12733} \frac{M^9}{r^9}\right.  \nonumber \\ 
& &  \left.  - \frac{433800}{1819} \frac{M^{10}}{r^{10}} - \frac{3483648}{1819} \frac{M^{11}}{r^{11}}\right) -  \frac{701429}{23708160} \frac{M^5}{r^5} \left( 1 + \frac{1013451}{701429} \frac{M}{r} \right. 
\nonumber \\ 
& & \left.  + \frac{1154835}{701429} \frac{M^2}{r^2} - \frac{3346744}{701429} \frac{M^3}{r^3} + \frac{3992148}{701429} \frac{M^4}{r^4} - \frac{9591516}{701429} \frac{M^5}{r^5}\right.  \nonumber \\  
& & \left.  + \frac{94091244}{701429} \frac{M^6}{r^6} - \frac{103967604}{701429} \frac{M^7}{r^7} - \frac{41345640}{701429} \frac{M^8}{r^8} \right. \nonumber \\  
& & \left. + \frac{109734912}{701429} \frac{M^9}{r^9}\right)  \left( 35 \cos ^4(\theta )-30 \cos ^2(\theta )+3 \right) \Bigg] \,, \\
g_{rr}^\dCS &=& \zeta \chi^4  \Bigg[ \frac{5}{384} \frac{M^4}{r^4 f\left(r\right)^3}  \left(1+\frac{577 }{175} \frac{M}{r}+\frac{8113}{1200} \frac{M^2}{r^2}-\frac{109309}{11760} \frac{M^3}{r^3} \right.  \nonumber \\ 
& & \left. +\frac{2125311}{3920} \frac{M^4}{r^4} +\frac{267403}{1470 } \frac{M^5}{r^5} + \frac{2001821}{420} \frac{M^6}{r^6} - \frac{19927289}{980} \frac{M^7}{r^7}\right.  \nonumber \\ 
& & \left. +\frac{22161021}{980} \frac{M^8}{r^8} -\frac{12553726}{245} \frac{M^9}{r^9} + \frac{249993}{5} \frac{M^{10}}{r^{10}} + \frac{735264}{25} \frac{M^{11}}{r^{11}}  \right) \nonumber \\ 
& & - \frac{1819}{56448} \frac{M^3}{r^3 f\left(r\right)^2 } \left(3 \cos ^2(\theta )-1\right)  \left(1+\frac{1084}{1819} \frac{M}{r}+\frac{62621}{12733 } \frac{M^2}{r^2}+\frac{56726}{1819} \frac{M^3}{r^3} \right.  \nonumber \\ 
& & \left. - \frac{2116475}{38199 } \frac{M^4}{r^4} - \frac{112168723}{381990} \frac{M^5}{r^5} - \frac{36858343}{190995} \frac{M^6}{r^6} + \frac{3546621}{9095} \frac{M^7}{r^7}\right.  \nonumber \\ 
& & \left. - \frac{47131846}{63665} \frac{M^8}{r^8} -\frac{5777844 }{12733 } \frac{M^9}{r^9} - \frac{32693976}{1819 } \frac{M^{10}}{r^{10}} + \frac{80123904}{1819 } \frac{M^{11}}{r^{11}}  \right) \nonumber \\ 
& & -  \frac{94699}{4741632} \frac{M^5}{r^5 f\left(r\right)} \left( 35 \cos ^4(\theta )-30 \cos ^2(\theta )+3 \right)  \left( 1 - \frac{916004}{473495 } \frac{M}{r} \right. 
\nonumber \\ 
& & \left.  + \frac{2411573}{473495} \frac{M^2}{r^2} + \frac{16109646}{473495} \frac{M^3}{r^3} + \frac{2585472}{43045} \frac{M^4}{r^4} - \frac{23898480}{94699} \frac{M^5}{r^5}\right.  \nonumber \\  
& & \left.  - \frac{314374068}{473495} \frac{M^6}{r^6} - \frac{41960268}{43045} \frac{M^7}{r^7} - \frac{1261951488}{473495 } \frac{M^8}{r^8} \right)  \Bigg] \,, \\
g_{\theta\theta}^\dCS &=& \zeta \chi^4  \Bigg[ \frac{67}{2240} \frac{M^5}{r^4}  \left(1+\frac{104533}{19296} \frac{M}{r}+\frac{583357}{45024} \frac{M^2}{r^2}+\frac{311763}{7504} \frac{M^3}{r^3} \right.  \nonumber \\ 
& & \left. +\frac{3112171}{33768} \frac{M^4}{r^4} +\frac{24899}{1608} \frac{M^5}{r^5} - \frac{2538845}{11256} \frac{M^6}{r^6} - \frac{190101}{268} \frac{M^7}{r^7}\right.  \nonumber \\ 
& & \left. -\frac{18648}{67} \frac{M^8}{r^8} \right) - \frac{1819}{56448} \frac{M^3}{r} \left(3 \cos ^2(\theta )-1\right) \left(1+\frac{17455}{7276}  \frac{M}{r}  \right.  \nonumber \\ 
& &  \left. +\frac{148755}{25466} \frac{M^2}{r^2} +\frac{52999}{3638 } \frac{M^3}{r^3} + \frac{3438929}{76398} \frac{M^4}{r^4}+ \frac{5163387}{63665} \frac{M^5}{r^5}\right.  \nonumber \\ 
& &  \left.  + \frac{14491811}{190995} \frac{M^6}{r^6} - \frac{5632}{85} \frac{M^7}{r^7} + \frac{6094488}{12733} \frac{M^8}{r^8} + \frac{1232136}{1819} \frac{M^9}{r^9}\right.  \nonumber \\ 
& &  \left.  + \frac{3483648}{1819} \frac{M^{10}}{r^{10}} \right) -  \frac{94699}{4741632} \frac{M^5}{r^3} \left( 1 + \frac{2984191}{1420485 } \frac{M}{r} \right. 
\nonumber \\ 
& & \left.  + \frac{2339824}{473495} \frac{M^2}{r^2} + \frac{94116}{8609} \frac{M^3}{r^3} + \frac{45539276}{1420485} \frac{M^4}{r^4} + \frac{16610916}{473495} \frac{M^5}{r^5}\right.  \nonumber \\  
& & \left.  + \frac{7853220}{94699} \frac{M^6}{r^6} - \frac{31810968}{473495} \frac{M^7}{r^7} - \frac{109734912}{473495} \frac{M^8}{r^8} \right) \nonumber \\  
& & \left( 35 \cos ^4(\theta )-30 \cos ^2(\theta )+3 \right) \Bigg] \,, \\
g_{\phi\phi}^\dCS &=& \zeta \chi^4  \Bigg[ \frac{1819}{211680} \frac{M^3}{r}  \left(1+\frac{17455}{7276} \frac{M}{r}+\frac{106545}{25466} \frac{M^2}{r^2}+\frac{571331}{58208} \frac{M^3}{r^3} \right.  \nonumber \\ 
& & \left. +\frac{47090579}{1222368} \frac{M^4}{r^4} +\frac{134570577}{1018640} \frac{M^5}{r^5} + \frac{315848443}{1527960} \frac{M^6}{r^6} \right.  \nonumber \\ 
& & \left. - \frac{118918819}{509320} \frac{M^7}{r^7} -\frac{52589025}{101864} \frac{M^8}{r^8} -\frac{11692683}{7276} \frac{M^9}{r^9}  + \frac{2308824}{1819} \frac{M^{10}}{r^{10}} \right)  \nonumber \\ 
& &  - \frac{9095}{592704 } \frac{M^3}{r} \left(3 \cos ^2(\theta )-1\right) \left(1+\frac{17455}{7276}  \frac{M}{r}  +\frac{740737}{272850} \frac{M^2}{r^2} \right. \nonumber \\ 
& & \left. + \frac{50790941}{13096800} \frac{M^3}{r^3}  + \frac{100159603}{4365600} \frac{M^4}{r^4}+ \frac{47638909}{727600} \frac{M^5}{r^5}\right.  \nonumber \\ 
& &  \left.  + \frac{30022409}{654840} \frac{M^6}{r^6} - \frac{84837063}{363800} \frac{M^7}{r^7} - \frac{5835501}{72760} \frac{M^8}{r^8} - \frac{29787597}{181900} \frac{M^9}{r^9}\right.  \nonumber \\ 
& &  \left.  + \frac{89318376}{45475} \frac{M^{10}}{r^{10}} \right) + \frac{1819}{658560} \frac{M^3}{r} \left( 1 + \frac{17455}{7276} \frac{M}{r} - \frac{198514}{180081} \frac{M^2}{r^2}  \right. 
\nonumber \\ 
& & \left.   - \frac{4120646}{540243} \frac{M^3}{r^3} - \frac{1309801}{360162} \frac{M^4}{r^4} + \frac{10009}{2805} \frac{M^5}{r^5}\right. - \frac{294356903}{2701215} \frac{M^6}{r^6}  \nonumber \\  
& & \left.  - \frac{109195222}{300135} \frac{M^7}{r^7} - \frac{4348890}{20009} \frac{M^8}{r^8} +\frac{7512372}{20009} \frac{M^9}{r^9} +\frac{54867456}{20009} \frac{M^{10}}{r^{10}}  \right) \nonumber \\  
& & \left( 35 \cos ^4(\theta )-30 \cos ^2(\theta )+3 \right)  + \frac{701429}{156473856} \frac{M^5}{r^3} \left( 1 + \frac{5962075}{2104287} \frac{M}{r} \right. \nonumber \\ 
& & \left.  + \frac{4434376}{701429} \frac{M^2}{r^2}   + \frac{7777884}{701429} \frac{M^3}{r^3} + \frac{59811476}{2104287 } \frac{M^4}{r^4} + \frac{28251588}{701429} \frac{M^5}{r^5}\right.  \nonumber \\  
& & \left. + \frac{66282516}{701429} \frac{M^6}{r^6} + \frac{4767336}{701429} \frac{M^7}{r^7} - \frac{109734912}{701429} \frac{M^8}{r^8} \right) \nonumber \\ & & \left(231 \cos ^6(\theta )-315 \cos ^4(\theta )+105 \cos ^2(\theta )-5 \right) \Bigg] \,, \\
g_{t\phi}^\dCS &=&  \zeta \chi^3 \sin ^2(\theta )  \Bigg[ -\frac{8819}{141120} \frac{M^4}{r^3}  \left(1+\frac{60155}{35276} \frac{M}{r}+\frac{8545}{8819} \frac{M^2}{r^2}-\frac{19828}{26457} \frac{M^3}{r^3} \right.  \nonumber \\ 
& & \left. +\frac{563669}{26457} \frac{M^4}{r^4} +\frac{549630}{8819} \frac{M^5}{r^5} + \frac{873180}{8819} \frac{M^6}{r^6} \right)  \nonumber \\ 
& &  - \frac{8819}{56448} \frac{M^4}{r^3} \left(3 \cos ^2(\theta )-1\right) \left(1+\frac{24875}{35276}  \frac{M}{r}  +\frac{95}{17638} \frac{M^2}{r^2} \right. \nonumber \\ 
& & \left. + \frac{90188}{26457} \frac{M^3}{r^3}  + \frac{684818}{26457} \frac{M^4}{r^4}+ \frac{385542}{8819} \frac{M^5}{r^5} + \frac{418572}{8819} \frac{M^6}{r^6} \right.  \nonumber \\ 
& &  \left. - \frac{508032}{8819} \frac{M^7}{r^7} \right) \nonumber \\ 
& & + \zeta \chi^5   \sin ^2(\theta ) \Bigg[ \frac{3840911}{142248960} \frac{M^4}{r^3}  \left(1+\frac{3368875}{7681822} \frac{M}{r}+\frac{539981961}{211250105} \frac{M^2}{r^2} \right.  \nonumber \\ 
& & \left. +\frac{63963088}{211250105} \frac{M^3}{r^3}  - \frac{28203665}{84500042 } \frac{M^4}{r^4} -\frac{218979789}{84500042} \frac{M^5}{r^5}  \right.  \nonumber \\ 
& & \left. + \frac{6554146711}{42250021} \frac{M^6}{r^6}  + \frac{1870270010}{3840911} \frac{M^7}{r^7} + \frac{3798260802}{3840911} \frac{M^8}{r^8} \right. \nonumber \\ 
& &  \left.  -\frac{1514962386}{3840911} \frac{M^9}{r^9}  - \frac{2505947220}{3840911} \frac{M^{10}}{r^{10}}  - \frac{1184222592}{3840911} \frac{M^{11}}{r^{11}} \right)  \nonumber \\ 
& &  + \frac{3840911}{56899584} \frac{M^4}{r^3} \left(3 \cos ^2(\theta )-1\right) \left(1+\frac{10036795}{7681822}  \frac{M}{r}  +\frac{949643961}{211250105} \frac{M^2}{r^2} \right. \nonumber \\ 
& & \left. + \frac{1196741284}{211250105} \frac{M^3}{r^3}  + \frac{304195064}{42250021} \frac{M^4}{r^4}+ \frac{646950168}{211250105} \frac{M^5}{r^5}\right.  \nonumber \\ 
& &  \left.  + \frac{19300145456}{211250105} \frac{M^6}{r^6} + \frac{1091987984}{3840911} \frac{M^7}{r^7} + \frac{13626382752}{19204555} \frac{M^8}{r^8} \right.  \nonumber \\ 
& &  \left. - \frac{914366016}{3840911} \frac{M^9}{r^9}  -  \frac{290957184}{3840911} \frac{M^{10}}{r^{10}}  -  \frac{3511517184}{3840911} \frac{M^{11}}{r^{11}}  \right) \nonumber \\ 
& & + \frac{65029949}{1738598400} \frac{M^6}{r^5} \left( 35 \cos ^4(\theta )-30 \cos ^2(\theta )+3 \right) \left( 1 + \frac{247489546}{195089847} \frac{M}{r} \right. 
\nonumber \\ 
& & \left.  - \frac{192857740}{585269541} \frac{M^2}{r^2}  + \frac{201416960}{195089847} \frac{M^3}{r^3} + \frac{6952033840}{195089847} \frac{M^4}{r^4} \right.  \nonumber \\  
& & \left. + \frac{49673623120}{585269541} \frac{M^5}{r^5} + \frac{8477276720}{65029949} \frac{M^6}{r^6}  - \frac{7984872720}{65029949} \frac{M^7}{r^7}  \right.  \nonumber \\ 
& & \left. - \frac{5714422560}{65029949 } \frac{M^8}{r^8} +\frac{8047226880}{65029949} \frac{M^9}{r^9}   \right) \Bigg] \,.
\end{eqnarray}

%%%%%%%%%%%%%%%%%%%%%%%%%%%%%%%%%%%%%%%%%%%%
\section*{References}

\bibliography{References}
\bibliographystyle{iopart-num}

\end{document}